\documentclass{article}
\usepackage{multirow}
\usepackage{amsmath}
\usepackage{arydshln}
\usepackage{graphicx}
\usepackage{threeparttable}
\usepackage{wrapfig}
\usepackage[preprint]{neurips_2026}

\makeatletter
\renewcommand{\@notice}{}
\makeatother
\usepackage{pgfplots}
\pgfplotsset{compat=1.18}
\usepackage{subcaption}
\usepackage{xcolor}
\definecolor{humancol}{RGB}{234,205,205}
\definecolor{aicol}{RGB}{224,181,173}
\definecolor{cocreatecol}{RGB}{215,207,188}
\definecolor{transformcol}{RGB}{183,198,214}
\usepackage{titletoc} 
\usepackage{hyperref} 
\usepackage[table]{xcolor}
\usepackage{booktabs}
\usepackage{makecell}
\usepackage{pgf}
\usepackage{graphicx}
\usepackage{nicematrix}
\definecolor{softblue}{RGB}{198,218,255}
\newcommand{\LOMOCell}[3]{\cellcolor{softblue!#1}\makecell{#2\\#3}}
\newcommand{\LOMOCellB}[3]{\cellcolor{softblue!#1}\makecell{\textbf{#2}\\\textbf{#3}}}

\usepackage{pgfplots}
\usepackage{tikz}
\usepgfplotslibrary{polar}
\pgfplotsset{compat=1.18}
\definecolor{humancol}{RGB}{234,205,205}
\definecolor{aicol}{RGB}{224,181,173}
\definecolor{cocreatecol}{RGB}{215,207,188}
\definecolor{transformcol}{RGB}{183,198,214}
\usepackage[utf8]{inputenc} % allow utf-8 input
\usepackage[T1]{fontenc}    % use 8-bit T1 fonts
\usepackage{hyperref}       % hyperlinks
\usepackage{url}            % simple URL typesetting
\usepackage{booktabs}       % professional-quality tables
\usepackage{amsfonts}       % blackboard math symbols
\usepackage{nicefrac}       % compact symbols for 1/2, etc.
\usepackage{microtype}      % microtypography
\usepackage{xcolor}         % colors
\usepackage{amsmath, amssymb, amsfonts, mathtools, bm, amsthm}
\usepackage{nccmath}   % for compact equations

\usepackage{xurl}
\usepackage{booktabs}   % professional tables
\usepackage{multirow}
\usepackage{makecell}
\usepackage{tabularx}
\usepackage{array}
\usepackage{colortbl}   % table cell coloring
\usepackage{subcaption}
\usepackage{graphicx}
\usepackage{tikz}
\usepackage{overpic}
\usepackage{pgfplots}
\pgfplotsset{compat=1.17}
\usepackage{soul}
\usepackage{comment}
\usepackage{nicematrix,tikz}
\usepackage{xcolor}
\usepackage{soul}       % \hl{} highlighting
\usepackage{gensymb}    % \degree symbol, etc.

\usepackage{algorithm}
\usepackage{algpseudocode}  % cleaner than algorithmic

\usepackage{xspace}
\usepackage{verbatim}
\usepackage{comment}
\usepackage{enumitem}
\usepackage{accents}
\usepackage{url}
\usepackage{pifont}
\usepackage[normalem]{ulem} % provides \uline, \uwave, etc.

\usepackage{graphicx}   % for \resizebox
\usepackage{booktabs}
\usepackage[table]{xcolor}
\usepackage{tabularx}
\usepackage{array}
\newcolumntype{C}{>{\centering\arraybackslash}X} % centered X column
\definecolor{darkblue}{RGB}{0,51,102}
\definecolor{lightblue}{RGB}{230,240,255}
\newcommand{\cmark}{\checkmark}
\newcommand{\xmark}{\ding{55}}

\usepackage{balance}

\usepackage[most]{tcolorbox}
\tcbset{
  mytakeawaybox/.style={
    colback=gray!10,
    colframe=gray!50,
    boxrule=0.5pt,
    arc=2mm,
    left=2mm, right=2mm,
    top=1mm, bottom=1mm,
    fonttitle=\bfseries,
    title={Takeaways}
  }
}
\newcommand{\BenchName}{\textsc{MuseBench}\xspace}

\title{\BenchName: A Benchmark for AI-Generated Music Detection and Artifact Analysis}

\renewcommand{\thefootnote}{\fnsymbol{footnote}}

\author{%
  Yi Chen Liu\thanks{Equal contribution.} \\
  University of Georgia \\
  \And
  Syed Irfan Ali Meerza\footnotemark[1] \\
  College of Engineering \\
  Virginia Commonwealth University \\
  \And
  Jiawei Yu\footnotemark[1] \\
  University of Georgia \\
  \AND
  Xi Guo \\
  University of Tennessee, Knoxville \\
  \And
  Samuel Bright \\
  Cornell University \\
  \And
  Jian Liu \\
  University of Georgia \\
}

\begin{document}
\maketitle
\renewcommand{\thefootnote}{\arabic{footnote}}
\setcounter{footnote}{0}

\begin{abstract}
Advances in generative AI have made it possible to produce music that is nearly indistinguishable from human compositions, raising the question of whether such works can be reliably distinguished from human compositions.
%Recent advances in generative AI have enabled models to produce music nearly indistinguishable from human compositions, raising questions about how humans and machines perceive and differentiate AI-generated works. 
We present \BenchName, a large-scale benchmark for AI-generated music detection and artifact analysis, comprising over 361K tracks. The dataset includes human compositions, stylistically matched music generated by ten state-of-the-art open-source and commercial models, such as Suno and Udio, as well as human-AI co-creations and variants processed with adversarial perturbations, compression, and acoustic post-processing.
%This study presents the first comprehensive systematic examination of human and AI capabilities in detecting and interpreting AI-generated music across a large benchmark of over 352K tracks, curated and released in this work as \BenchName. The dataset encompasses authentic human compositions and stylistically matched pieces generated by ten state-of-the-art open-source and commercial models (e.g., Suno, Udio), as well as human-AI co-creations and variants subjected to adversarial perturbations, compression, and acoustic post-processing, etc. 
%We benchmark ten automated detectors to evaluate detection accuracy, robustness, and perceptual strategies. Through the first comprehensive analysis combining human judgments with acoustic feature inspection, we uncover recurring artifacts, such as non-idiomatic harmonic progressions, upsampling-induced spectral periodicities, and violations of physical performance constraints, that shape both human and machine perception. Our findings highlight complementary strengths and vulnerabilities of human and machine listeners, providing new insights into perceptual boundaries and guiding the design of more reliable music authenticity detection systems.
Using \BenchName, we benchmark 11 automated detectors
and evaluate their accuracy, generalization, and robustness, under default, out-of-domain, and transformed-variant experiment settings. We further conduct a comprehensive artifact analysis through acoustic feature inspection, identifying recurring signatures such as non-idiomatic harmonic progressions, upsampling-induced spectral periodicities, and violations of physical performance constraints. Our findings reveal key strengths and vulnerabilities of existing detectors and provide practical insights for building more reliable AI music detection systems.
\end{abstract}
\vspace{-2mm}

\section{Introduction}
\vspace{-2mm}
\label{Sec:Introduction}
Recent advances in generative AI have made high-fidelity music generation broadly accessible. Models such as Stable Audio Open~\cite{evans2025stable} and Riffusion~\cite{Forsgren_Martiros_2022}, together with commercial platforms such as Suno~\cite{sunoai2023} and Udio~\cite{udio2024}, can now produce realistic songs from simple text prompts. While these systems expand creative possibilities, they also create urgent challenges for music authenticity~\cite{recordingacademyArtificialIntelligence}, platform integrity~\cite{nytimesAIFake, musicbusinessworldwideItsTime}, and generative-AI governance, particularly around the principles of consent, credit, and compensation~\cite{, spotifyRoyalties, gervais2015collective, sunoRevenue}. As AI-generated tracks increasingly enter streaming catalogs and open-upload platforms, reliable detection and artifact analysis have become essential for distinguishing synthetic music from human-created works.

Existing AI-music detection resources remain limited in scope. Prior benchmarks (e.g. M6~\cite{li2024m6}, FakeMusicCaps~\cite{comanducci2024fakemusiccaps}) typically focus on a small number of generators, emphasize narrow categories such as vocal music (SONICS~\cite{rahman2025sonics}), and evaluate detectors primarily under clean, controlled conditions.
However, real-world music pipelines involve diverse generators, human-AI co-creation, lossy compression, acoustic effects, format conversion, and adversarial or additive perturbations. 
In parallel, a growing line of work has focused on identifying artifacts in AI-generated music, uncovering signals such as spectral upsampling patterns~\cite{afchar2025fourier} and phase incoherence~\cite{afchar2025challenges}, but the field still lacks a unified benchmark that connects large-scale detector evaluation with systematic analysis of recurring musical, spectral, vocal, and performance-related artifacts.

To address these gaps, we present \BenchName, a large-scale benchmark for AI-generated music detection and artifact analysis. \BenchName contains 361,166 tracks ($\approx$5,296 hours), including verified human compositions, outputs from ten state-of-the-art text-to-music generators, human-AI co-created tracks, transformed variants produced through lossy compression, acoustic post-processing, adversarial perturbations, and additive noise, and an out-of-domain dataset. Compared with existing benchmarks, \BenchName is broader in generator coverage, more realistic in its inclusion of post-processing and perturbation conditions, and richer in its support for artifact-level analysis across both instrumental and vocal music.

%Using \BenchName, we evaluate eleven automated detectors, including CNN-based, Transformer-based, waveform-based models, and two commercial APIs. 
Using \BenchName, we evaluate 11 automated detectors spanning diverse architectures and input representations, including handcrafted spectral artifact features with statistical models, CNN- and Transformer-based models, waveform- and spectrogram-based approaches, as well as two commercial APIs.
Beyond accuracy, we examine robustness across generators, genres, vocals, co-created content, and transformed variants. This evaluation reveals where current detectors generalize, where they fail, and how model design choices influence sensitivity to different AI-music artifacts.
We further conduct a systematic artifact analysis of AI-generated music. Our analysis spans frequency-band sensitivity, spectral fingerprints, harmonic irregularities, physical performance-constraint violations, and vocal artifacts. These results provide a structured taxonomy of recurring generative artifacts and offer practical guidance for designing more robust and interpretable AI music detection systems. Our main contributions are as follows:

\begin{itemize}[leftmargin=*, itemsep=0pt, parsep=2pt, topsep=2pt, partopsep=0pt]
\item \textbf{\BenchName: a large-scale benchmark for AI-music detection.} We curate 361,166 tracks spanning verified human compositions, tracks generated by 10 state-of-the-art generators, human-AI co-created music, and transformed variants obtained by applying compression, acoustic effects, adversarial perturbations, and additive noise. To our knowledge, \BenchName is the first benchmark to jointly cover generator diversity, co-creation, and realistic transformations at this scale.

\item \textbf{Comprehensive evaluation of automated detectors.} We benchmark 11 detectors, including handcrafted feature-based, CNN-based, Transformer-based, and two commercial APIs, evaluating their detection accuracy and failure patterns under default, out-of-domain, and transformed-variant experiment settings.
% , co-created tracks, and post-processing transformations.

\item \textbf{Systematic artifact analysis of AI-generated music.} We identify recurring spectral, harmonic, vocal, and performance-related artifacts through frequency-sensitivity profiling, spectral analysis, chord-grammar inspection, physical performance-constraint analysis, and vocal artifact assessment.

\item \textbf{Design insights for robust AI-music detection.} We derive practical recommendations for future detectors, including artifact-aware augmentation, balanced spectral sensitivity, and incorporation of structural, vocal, and performance-informed cues.
\end{itemize}

\vspace{-4mm}

\section{Related Work}
\vspace{-2mm}
\label{Sec:Related Work}
%\textbf{AI Music Generators.} Computer-generated music traces back to Hiller’s ILLIAC Suite~\cite{hiller1958musical}, with neural approaches emerging in the late 1980s~\cite{nierhaus2009algorithmic}. Modern deep generative models have since transformed the field. Early systems such as MuseVAE~\cite{roberts2018hierarchical}, MuseGAN~\cite{dong2018musegan}, and Jukebox~\cite{dhariwal2020jukebox} explored long-term structure, multi-track composition, and high-fidelity synthesis. SymphonyNet~\cite{liu2022symphony} further advanced orchestral modeling, paving the way for today’s text-conditioned frameworks. Recent models emphasize controllability and natural-language conditioning. Mubert~\cite{mubeert} uses tag- and prompt-based control; Riffusion~\cite{Forsgren_Martiros_2022} adapts Stable Diffusion~\cite{rombach2022high} to spectrograms; and MusicLM~\cite{agostinelli2023musiclm} and MusicGen~\cite{copet2023simple} enable high-fidelity, text-aligned, long-form generation. Diffusion-based systems, including Jasco~\cite{tal2024joint}, MAGNeT~\cite{ziv2024masked}, AudioLDM2~\cite{liu2024audioldm2}, Stable Audio~\cite{evans2025stable}, and Mustango~\cite{melechovsky2024mustango}, expand controllability to rhythm, harmony, and editing tasks. Lyria~\cite{lyriateam2025livemusicmodels} instead adopts a real-time codec language model over SpectroStream tokens for low-latency generation. Commercial platforms like Suno~\cite{sunoai2023} and Udio~\cite{udio2024} have popularized these advances, making it possible for non-experts to create full-length vocal tracks within seconds. 
\textbf{AI Music Generators.}
Recent advances in deep generative models have enabled high-fidelity, controllable music generation. Early neural systems such as MuseVAE~\cite{roberts2018hierarchical}, MuseGAN~\cite{dong2018musegan}, and Jukebox~\cite{dhariwal2020jukebox} explored hierarchical composition, multi-track generation, and high-fidelity synthesis. Modern text-conditioned models, including MusicLM~\cite{agostinelli2023musiclm}, MusicGen~\cite{copet2023simple}, AudioLDM2~\cite{liu2024audioldm2}, Stable Audio Open~\cite{evans2025stable}, MAGNeT~\cite{ziv2024masked}, Jasco~\cite{tal2024joint}, and Mustango~\cite{melechovsky2024mustango}, further improve controllability over style, rhythm, harmony, and editing. Commercial platforms such as Suno~\cite{sunoai2023} and Udio~\cite{udio2024} have made these capabilities widely accessible, enabling non-experts to generate full-length songs from simple prompts. This rapid progress increases the diversity and realism of AI-generated music, creating new challenges for detection and evaluation.

\textbf{Automated AI-Music Detection.}
Automated detection methods broadly fall into three categories: statistical approaches using hand-crafted audio features and classical models such as Random Forest and SVM~\cite{singh2021qsvm}; CNN-based audio classifiers, including RawNet2~\cite{tak2021rawnet2}, ResNet variants~\cite{he2016deep}, ConvNeXt~\cite{liu2022convnet}, and M5~\cite{dai2017very}; and Transformer-based models such as ViT~\cite{dosovitskiy2021vit}, AST~\cite{gong2021ast}, and SpecTTTra~\cite{rahman2025sonics}. A complementary line of work studies artifact-based forensics. Afchar \textit{et al.}~\cite{afchar2024detecting,afchar2025challenges} analyze spectrogram-based cues for detecting synthetic audio and further identify deconvolution-induced spectral peaks through Fourier-domain analysis~\cite{afchar2025fourier}. Other signals, such as lyric-based cues, have also been explored~\cite{frohmann2025ai}. Commercial systems, including Deezer's internal AI-tagging system~\cite{newsroomdeezerDeezerDeploys}, Believe's AI Radar~\cite{musicbusinessworldwideBelieveDeveloped}, Audible Magic's Version ID~\cite{audiblemagicIdentifyingCover}, Ircam Amplify~\cite{ircamamplifyMusicDetector}, and Resemble AI~\cite{resembleDetectDeepfakes}, further highlight practical demand. However, many commercial systems are proprietary, limiting reproducible evaluation.

\textbf{Benchmarking AI-Music Detection.}
Recent benchmarks have begun to evaluate AI-music detection using curated datasets and standardized tasks. FakeMusicCaps~\cite{comanducci2024fakemusiccaps} provides an early corpus for text-to-music detection, with approximately 27K tracks from several generator families, but its scale and detector coverage remain limited. SONICS~\cite{rahman2025sonics} expands the benchmark to approximately 97K songs and includes commercial generators such as Suno and Udio, but focuses primarily on vocal music, which may bias detection toward linguistic or pronunciation artifacts and limit generalization to instrumental tracks. M6~\cite{li2024m6,li2024xai} further expands coverage across generators, genres, and instruments, but does not fully capture practical workflows such as codec/transcoding diversity, human-AI collaboration, per-stem analysis, or robustness under perturbation-based evasion.

%Overall, existing benchmarks remain limited in three key dimensions: limited generator diversity, insufficient coverage of realistic post-processing and evasion conditions, and lack of systematic artifact-level analysis linking detector behavior to underlying generative signatures. Inspired by broader studies on AI-generated content detection under manipulations~\cite{ha2024organic} and human detection limitations across modalities~\cite{10646666}, we propose \BenchName, a large-scale, multi-dimensional benchmark that jointly evaluates detector performance across diverse generators, human-AI co-created content, and realistic transformations, while enabling systematic analysis of recurring artifacts in AI-generated music.

\vspace{-3mm}

\section{Benchmark Details}
\vspace{-3mm}
Our benchmark is constructed to support clean-provenance and reproducible evaluation across multiple interacting dimensions, including musical coverage, generator breadth, style-aligned prompting, detector diversity, and robustness to transformations and attacks. We defer the detailed dataset design and construction protocol to Appendix~\ref{sec: generation-pipeline}. \BenchName is publicly available at \url{https://huggingface.co/Anonymousv22222/datasets}.
\vspace{-2mm}

\subsection{Evaluating Machine Capabilities}
\vspace{-2mm}

\label{subsec:detectors}
For machine-based detectors, we evaluate a comprehensive suite of 11 state-of-the-art systems. To ensure a robust analysis across diverse inductive biases and input modalities, we select nine open-source research baselines that span diverse architectural paradigms. These include classical machine learning and theory-driven spectral analysis (\textbf{Q\text{-}SVM}~\cite{singh2021qsvm}, \textbf{Deezer}~\cite{afchar2025fourier}\footnote{This theory-driven fingerprint detector leverages spectral peaks from deconvolution artifacts and has been deployed in Deezer’s AI music detector.~\cite{afchar2025fourier}.}) to raw waveform convolutional networks (\textbf{M5}~\cite{dai2017very}, \textbf{RawNet2}~\cite{tak2021rawnet2}), spectrogram-based CNNs (\textbf{ConvNeXt}~\cite{liu2022convnet}, \textbf{ResNet18+Spec}~\cite{he2016deep}), and global self-attention transformers-based models (\textbf{ViT}~\cite{dosovitskiy2021vit}, \textbf{AST}~\cite{gong2021ast}, \textbf{SpecTTTra}~\cite{rahman2025sonics}). In addition to these open-weight models, we evaluate two proprietary commercial detectors (\textbf{Ircam Amplify}~\cite{ircamamplifyMusicDetector}, \textbf{Resemble AI}~\cite{resembleDetectDeepfakes}\footnote{While Resemble AI is primarily known for deepfake speech detection, it also provides support for AI-generated music detection~\cite{resemble_music}.}), accessed as black-box APIs to represent current industry capabilities. 
%\textbf{Deezer} introduces a theory-driven fingerprint detector using spectral peaks from deconvolution artifacts. 
%\textbf{Resemble AI} is a leader in deepfake speech detection~\cite{11345101}, which can also be applied to music detection. 
A detailed summary of all evaluated detectors, including their categories, backbone architectures, input representations, parameter counts, and training configurations, is provided in Appendix~\ref{sec: appendix-detectors}.
\vspace{-2mm}

\subsection{Overview \& Benchmark Comparison}
\vspace{-2mm}

In order to perform a comprehensive examination of AI capabilities in detecting and interpreting AI-generated music, a benchmark dataset must capture diverse and interacting dimensions of music generation, transformation, and perception, as outlined in Appendix Section~\ref{subsec:objectives_challenges}. However, existing benchmark datasets remain limited in scope and fail to represent these multidimensional factors. As shown in Appendix Table~\ref{tab:dataset_comparison}, 
M6~\cite{li2024m6} includes both vocal and instrumental tracks but covers only a few generation models and lacks diversity in content transformations. FakeMusicCaps~\cite{comanducci2024fakemusiccaps} expands generator coverage but still focuses on clean, unmodified samples, omitting transformed variants that often appear in real-world remixing and distribution. SONICS~\cite{rahman2025sonics} expands dataset scale and includes commercial generators, but focuses predominantly on vocal music and similarly lacks systematic treatment of content transformations.
%SONICS~\cite{rahman2025sonics} provides a larger collection of AI-generated songs but focuses primarily on vocal material and similarly lacks any systematic coverage of content modifications.
%Importantly, none of these datasets include human-AI co-creations, which are increasingly common in real-world music production and represent a critical middle ground between fully synthetic and fully human works. Likewise, none consider transformed variants, such as compressed, acoustically post-processed, or adversarially perturbed versions, that can substantially affect both human perception and model robustness.
More importantly, these benchmarks omit two key aspects of modern music production and distribution. First, none include human-AI co-created music, which is increasingly prevalent and occupies a critical middle ground between fully human and fully synthetic compositions. Second, they do not account for transformed variants such as compression, acoustic post-processing, or adversarial perturbations, which frequently arise in real-world pipelines and can significantly impact detector robustness.

\begin{sloppypar}

To bridge these gaps, we introduce \BenchName, a large-scale benchmark comprising over 361K tracks generated from 10 diverse music models. \BenchName brings together vocal, instrumental, and human-AI co-created music, systematically transformed variants produced through compression, acoustic post-processing, and adversarial perturbations, along with a fully out-of-domain human-music dataset. This design enables controlled and comprehensive evaluation of both perceptual and algorithmic robustness under realistic and adversarial conditions.
%together with a wide range of compressed, post-processed, and adversarially perturbed variants, enabling systematic analysis of perceptual and algorithmic robustness across realistic and adversarial conditions. 
Detailed statistics of \BenchName, including track counts, total duration, and music categories, are provided in Appendix Table~\ref{tab:datasetoverview}.
\end{sloppypar}
\vspace{-3mm}

\subsection{Dataset Construction}
\label{sec:dataset}
\begin{figure}[t]
  \centering
   \includegraphics[width=0.86\linewidth]{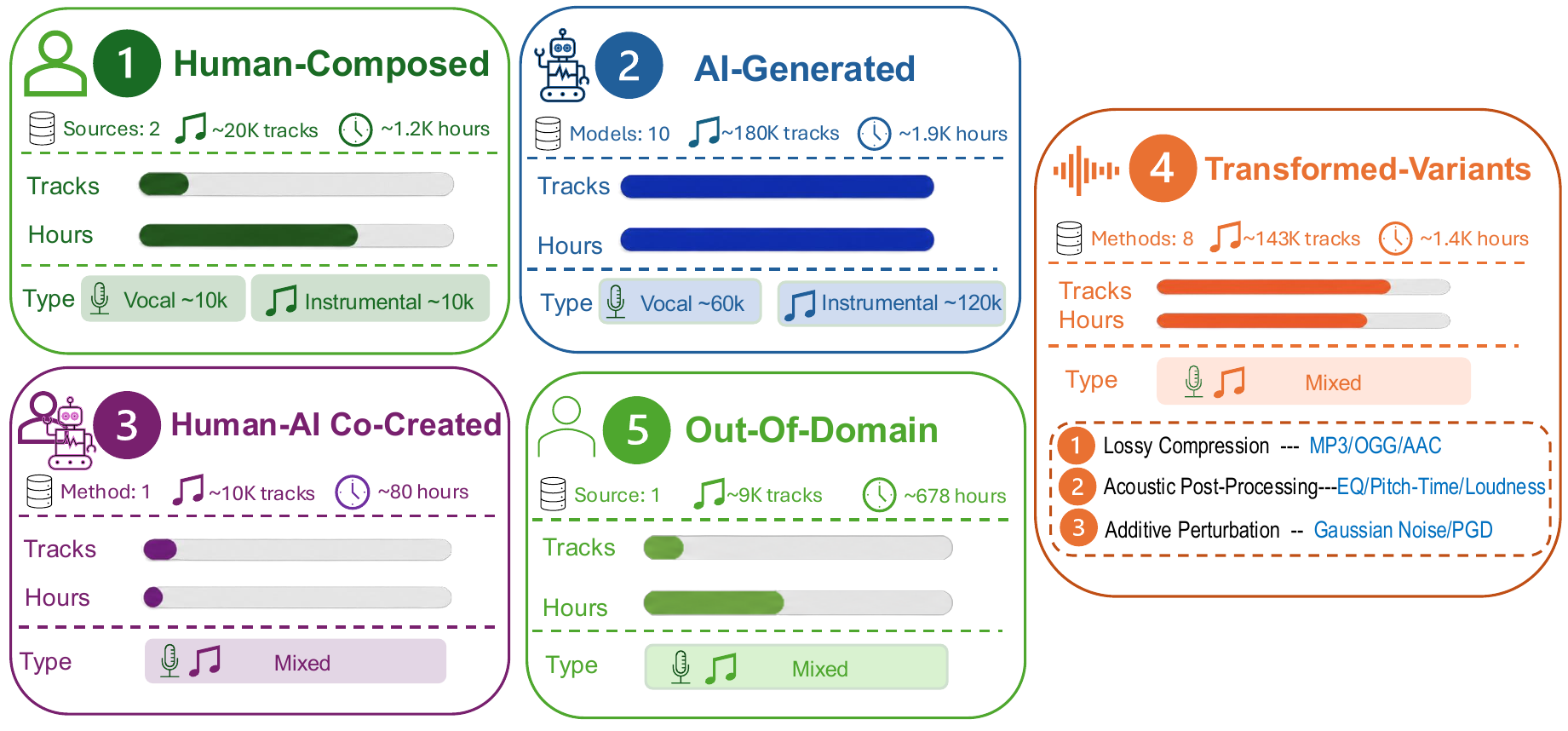}
  \caption{Overview of \BenchName.}
  \vspace{-6mm}
  \label{fig: dataset}
\end{figure}

\vspace{-3mm}
\subsubsection{\BenchName: Human-Composed Music}
\label{sec:human-composed}
\vspace{-3mm}
Human-composed music serves as the reference ground truth for evaluating machine ability to distinguish synthetic compositions. To ensure reliability, the dataset must provide broad coverage across both vocal and instrumental music, along with verifiable metadata, such as artist, album, and release information, to confirm human authorship.
%Human-composed music serves as the ground truth for evaluating both human and AI abilities to distinguish synthetic compositions. To ensure fairness and authenticity, the dataset must encompass both vocal and instrumental works for comprehensive coverage and include verifiable metadata, such as album title, artist name, and release year, to confirm their genuine human origin.
To this end, we construct the human subset using two complementary datasets. For vocal music, we use Genius-Lyrics~\cite{geniuslyrics}, which contains over 3 million tracks spanning diverse genres, including pop, hip-hop, and R\&B. For instrumental music, we adopt JamendoMaxCaps~\cite{roy2025jamendomaxcaps}, comprising more than 362K tracks with similarly broad genre coverage. Both datasets include rich metadata (e.g., artist and album information), enabling verification of their human origin. 
%We employ two music datasets that satisfy the aforementioned conditions. For vocal music, we use Genius-Lyrics~\cite{geniuslyrics}, which includes over 3 million vocal tracks from the Genius platform spanning diverse genres such as pop, hip-hop, and R\&B. For instrumental music, we adopt JamendoMaxCaps~\cite{roy2025jamendomaxcaps}, comprising more than 362K instrumental tracks from Jamendo, spanning diverse genres as well. Both datasets provide rich metadata, e.g., artist name, and album title, allowing verification of their human origin. 
We authenticated the music sources (Youtube and Jamendo) and its metadata for correctness, and randomly sampled metadata for 10,000 tracks from each dataset with each generator computationally feasible. After excluding unavailable entries, we obtained 9,983 Genius-Lyrics tracks and 9,920 JamendoMaxCaps tracks, totaling 19,903 verified human-composed.
\vspace{-2mm}
\subsubsection{\BenchName: AI-Generated Music}
\vspace{-2mm}
\label{sec:AI-Generated Music}
To fairly evaluate AI abilities in distinguishing synthetic music, the AI-generated benchmark must meet two requirements: broad generator coverage across open-source and commercial models, and stylistic alignment with human compositions. This alignment prevents genre, tempo, or production-quality differences from becoming superficial cues, enabling evaluation to focus on authenticity rather than trivial artifacts.
%To enable a fair and comprehensive evaluation of human and AI capabilities in distinguishing synthetic compositions, the AI-generated component of the benchmark must satisfy two key requirements.
%the AI-generated component of the benchmark must satisfy two key requirements.
%First, it should provide broad generator coverage, encompassing diverse state-of-the-art models across both open-source and commercial systems to reflect real-world generation diversity.
%Second, AI-generated tracks must be stylistically aligned to human compositions. Without such alignment, differences in genre, tempo, or production quality could create superficial cues, allowing models or listeners to rely on trivial stylistic artifacts rather than genuine perceptual or structural distinctions. Enforcing stylistic consistency isolates the authenticity dimension and enables meaningful comparison between human and AI creativity.

\noindent\textbf{Text-to-Music Generators.}
We include ten text-to-music generators spanning diverse architectures and accessibility levels to ensure representativeness, as shown in Appendix Table~\ref{tab:generatorsversion}, we include seven open-source models that reflect the trends in academic research: 
\textbf{MusicGen}~\cite{copet2023simple}, a Transformer-based autoregressive model for text-to-music generation; \textbf{AudioLDM2}~\cite{liu2024audioldm2}, a diffusion-based framework trained on large-scale audio–text pairs; \textbf{Riffusion}~\cite{Forsgren_Martiros_2022}, which leverages latent diffusion over spectrogram representations; \textbf{MAGNeT}~\cite{ziv2024masked}, a masked generative transformer enabling high-quality music synthesis; \textbf{JASCO}~\cite{tal2024joint}, a flow-matching model that jointly conditions on chords, drums, and melody; \textbf{Mustango}~\cite{melechovsky2024mustango}, a diffusion-based system emphasizing long-horizon coherence; and \textbf{Stable Audio Open}~\cite{evans2025stable}, an open diffusion model trained for high-fidelity audio generation. To complement these, we incorporate three closed-source commercial systems that represent the frontier of industrial deployment: \textbf{Lyria RealTime}~\cite{lyriateam2025livemusicmodels} from Google DeepMind, a Transformer model with real-time generation; \textbf{Suno}~\cite{sunoai2023}, offering end-to-end music creation with vocal and lyric conditioning; and \textbf{Udio}~\cite{udio2024}, a full-generation platform emphasizing vocal–instrumental coherence, for which we used two variants, Udio 32 and Udio 130, capable of generating songs of 32 and 130 seconds, respectively. Collectively, these ten generators cover the dominant paradigms of contemporary AI music creation, balancing academic reproducibility with real-world relevance.

\noindent\textbf{Music Generation Pipeline for Stylistic Alignment.}
To ensure that AI-generated music is directly comparable to human compositions, we design a controlled caption-based generation pipeline that maintains stylistic consistency across both categories (Appendix~\ref{sec: dataset-generation}).
\vspace{-2mm}
\subsubsection{\BenchName: Human-AI Co-Creations}
\label{sec: co-created}
\vspace{-2mm}
An increasing portion of modern music is neither purely human-composed nor fully AI-generated, but emerges from iterative human–AI collaboration. To capture this setting, we introduce a human-AI co-created category in \BenchName.
In \BenchName, we define human-AI co-created music as human-composed tracks that are subsequently remixed by AI generation models based on expert feedback derived from a smaller, professionally reviewed subset of compositions. As illustrated in Appendix Figure~\ref{fig:co-creations}, the process begins with a set of candidate human-composed tracks evaluated by a panel of $31$ professional musicians, each providing qualitative suggestions (e.g., enhancing harmonic richness, adjusting rhythmic dynamics, or adding atmospheric elements). Because collecting expert feedback at scale is infeasible, we use the OpenL3 model~\cite{cramer2019look} to extract perceptual features from the reviewed tracks and identify additional human compositions with similar timbral and structural characteristics. These matched tracks are then paired with the corresponding expert feedback. The original natural-language comments are subsequently refined and standardized using GPT-OSS~\cite{agarwal2025gpt}, producing consistent music-editing prompts.
%A growing share of music today is neither purely human-composed nor fully AI-generated but results from collaborative workflows where humans and machines iteratively refine a composition. To reflect this emerging paradigm, we include a human-AI co-created category in our benchmark dataset. 
%This addition bridges the gap between human and machine works, supporting studies of hybrid creativity and improving detection models that must recognize increasingly subtle human-AI collaborations.
%Each musician provided qualitative suggestions for improving the music, such as adding harmonic resonance, adjusting rhythmic dynamics, or introducing atmospheric layers.
%Because obtaining expert feedback for thousands of tracks was not feasible, we used the OpenL3 model~\cite{cramer2019look} to extract perceptual features from the reviewed pieces and identify other human-composed tracks with high similarity in timbre and structure. These perceptually similar tracks were then paired with the corresponding expert feedback. The natural-language comments from the musicians were refined and structured using GPT-OSS~\cite{agarwal2025gpt}, which transformed them into consistent and detailed music-editing prompts. An example of the original feedback and the resulting structured prompt is shown in Appendix Table~\ref{tab:co-created_example}. 
Finally, both the reviewed and matched tracks, along with their associated prompts, are input to a text-to-music generator (MusicGen-Style~\cite{rouard2024audio}) to produce 30-second remixed outputs at a 32~kHz sampling rate, which are then upsampled to 48~kHz. This pipeline enables scalable synthesis of human–AI co-created music while preserving the intent of expert feedback and extending it across a broader set of stylistically aligned compositions. More details about human-AI co-creations can be found in Appendix~\ref{sec:example_cocreated}.

\vspace{-2mm}
\subsubsection{\BenchName: Transformed Variants}
\label{sec: Perturbed}
\vspace{-2mm}
To simulate real-world storage, sharing, and evasion conditions, we apply three groups of practical transformations to AI-generated tracks in \BenchName. These transformations reflect common operations in music production and distribution pipelines, as well as intentional manipulations designed to obscure provenance. Studying their effects enables us to assess detector robustness under realistic and adversarial scenarios. We briefly describe each group below, with detailed parameters provided in Appendix~\ref{sec:additional_dataset}.

\noindent \textbf{Lossy Compression.} 
Lossy compression is ubiquitous in online streaming and file distribution, where storage and bandwidth efficiency take precedence over audio fidelity. We apply three widely adopted codecs, such as MP3, OGG Vorbis, and AAC, to evaluate how compression-induced artifacts and information loss affect detector performance across different perceptual encoding schemes.

\noindent\textbf{Acoustic Post-processing.} 
Acoustic post-processing is routinely used in professional and user-generated workflows for mastering, remixing, or loudness normalization. To capture this diversity, we consider three operations: equalization (EQ), pitch/time shifts, and loudness normalization. These represent typical studio or consumer adjustments that modify spectral balance, tempo, and dynamic range, providing insight into the stability of detectors under musically meaningful transformations.

\noindent\textbf{Additive Perturbations.}
In practical or adversarial settings, users may intentionally introduce perturbations to alter AI-generated content or evade detection. We therefore include two representative types: Gaussian noise, modeling natural recording imperfections or low-quality re-encodings, and adversarial perturbations~\cite{madry2017towards}, representing deliberately optimized changes crafted to mislead detectors. Together, these cases span both unintentional and strategic interference scenarios, enabling systematic robustness evaluation.

\vspace{-2mm}
\subsubsection{\BenchName: Out-Of-Domain Human-music Dataset}
\vspace{-2mm}
To further evaluate detectors trained on our benchmark using a fully Out-Of-Domain human-music dataset, we construct a human-only evaluation pool of 8,672 tracks from the Free Music Archive (FMA)~\cite{defferrard2016fma}. The pool is restricted to music uploaded before 2010 and includes both vocal and purely instrumental tracks, yielding a reliable collection of human-composed music that is temporally separated from recent AI-music generators.
\vspace{-2mm}
\vspace{-2mm}
\section{Experiments}
\label{Sec:Experiments}
\vspace{-2mm}
\subsection{Experimental Setup}
%In this section, we evaluate the machine-based detectors introduced in Section~\ref{subsec:detectors}, which include eight open-source models and two commercial systems. 
We evaluate the proposed benchmark across a diverse set of machine-based detectors introduced in Section~\ref{subsec:detectors}, and their implementation details are provided in Appendix~\ref{sec:detectors}.
To assess detection performance, we conduct evaluations on human-composed (negative class) versus AI-generated music (positive class) under two main experimental settings: 
\emph{Default} and \emph{Leave-One-Model-Out (LOMO)}. 
%In the \emph{Default} setting, the human and AI-generated subsets in \BenchName are split 70/20/10 into training, validation, and testing. 
%Unless otherwise specified, we randomly sample a 10-second excerpt from each track for inference. 
%Results using 30-second excerpts are reported in Appendix~\ref{subsec:30s}. 
In the \emph{Default} setting, we partition \BenchName into 70/20/10 splits for training, validation, and testing across both human and AI-generated samples. For inference, we randomly sample a 10-second excerpt from each track unless stated otherwise; results using longer 30-second excerpts are deferred to Appendix~\ref{subsec:30s}.
%In the \emph{LOMO} setting, motivated by prior findings~\cite{li2024m6}
%indicating strong in-domain yet weak out-of-domain generalization, 
%we hold out one AI generator for testing to assess cross-model robustness. 
%The remaining AI-generated data are split 80/20 into training and validation, while the human subset follows the same protocol as in the \emph{Default} setting.

To evaluate cross-model generalization, we adopt a \emph{Leave-One-Model-Out (LOMO)} protocol and an \emph{Out-Of-Domain (OOD)} protocol. For \emph{LOMO}, we hold out all samples from a single AI generator for testing, following prior observations~\cite{li2024m6} that detectors often exhibit strong in-domain but limited out-of-domain performance. The remaining AI-generated data are used for training and validation with an 80/20 split, while the human subset is kept consistent with the \emph{Default} setting. For \textit{OOD}, since the dataset contains only human music, we report the accuracy of trained detectors in the \emph{Default} setting in correctly classifying these samples as human.
Additionally, we examine the robustness of these detectors under \textit{human-AI co-creation}  and diverse \textit{transformations} workflows included in \BenchName. 
For this robustness analysis, we evaluate all open-source detectors using their best checkpoints from the \emph{Default} setting, and additionally include two commercial systems to provide a comparison with industry-grade, black-box detectors. Specifically, we consider \emph{Ircam Amplify} and \emph{Resemble AI}, both of which are widely used in practice.
%alongside the two commercial systems.
%\noindent\textbf{Commercial Detectors.} To benchmark industry-facing, black-box systems under the same evaluation protocol, we include two widely used commercial services: \emph{Ircam Amplify} and \emph{Resemble AI}. 
Due to the costs associated with API access, both detectors were applied to a representative subset of \BenchName, consisting of 1{,}990 human-composed tracks and 4{,}500 AI-generated tracks (including 250 from Suno and Udio, and 500 from each of the remaining generators) under the \emph{Default} setting. 
For robustness evaluation, we additionally consider 200 human-AI co-created tracks and 300 samples from each of the transformed variants in \BenchName (2{,}400 in total). 
Predictions from \emph{Ircam Amplify} include a confidence score, for which we use a decision threshold of 0.5, whereas \emph{Resemble AI} provides only categorical outputs (\emph{Real} or \emph{Fake}) without confidence estimates.
%\emph{Ircam Amplify} outputs a Boolean label accompanied by a confidence score, for which we adopt a threshold of 0.5, whereas \emph{Resemble AI} provides only a categorical output (\emph{Real} or \emph{Fake}) without confidence estimation.

\noindent\textbf{Evaluation Metrics.} To comprehensively assess detector performance, we report metrics capturing both accuracy and error behavior. Specifically, we use the following metrics:
\begin{itemize} [leftmargin=*, itemsep=0pt, topsep=0pt]
    \item \textbf{Balanced Accuracy (ACC$_B$):} The average of the true positive rate and true negative rate, ensuring that performance is not biased toward the majority class.
    \item \textbf{Precision:} The proportion of samples predicted as AI-generated that are correctly classified as AI-generated music.
    \item \textbf{Recall:} The proportion of AI-generated music that are correctly identified as AI-generated.
    \item \textbf{False Positive Rate (FPR):} The proportion of human compositions incorrectly classified as AI-generated music.
    \item \textbf{False Negative Rate (FNR):} The proportion of AI-generated music incorrectly classified as human samples.
\end{itemize}
\subsection{Overall Detector Performance}
\subsubsection{Overall Machine Capabilities}

\begin{wrapfigure}[17]{r}{0.5\linewidth}
  \centering
  \vspace{-12mm}
  \includegraphics[width=\linewidth]{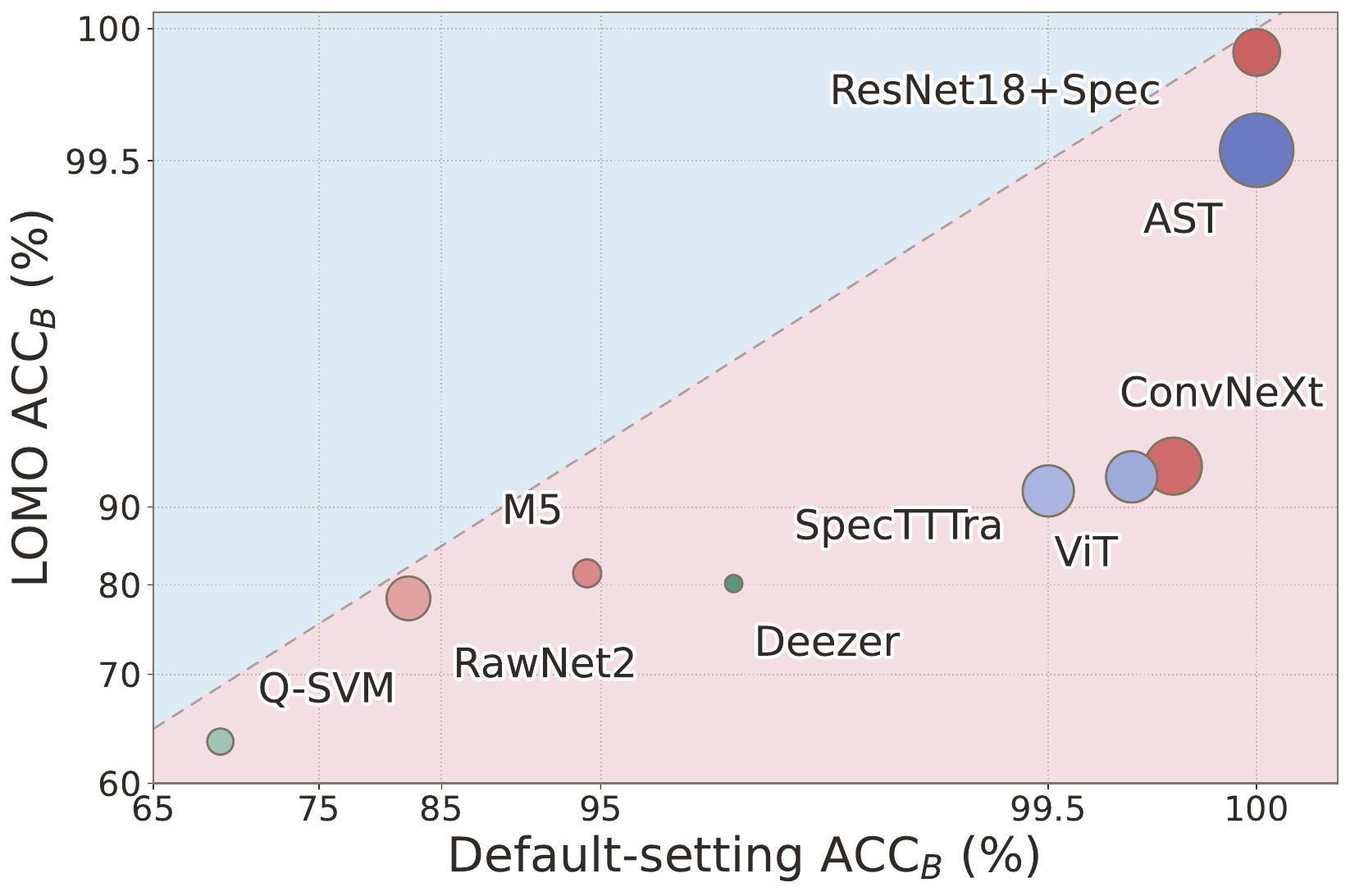}
  \vspace{-3mm}
  \caption{Detector performance under default and LOMO settings. Complete numerical results are reported in Appendix Table~\ref{tab:in_domain} and Table~\ref{tab:leave_one_out}. Bubble size is proportional to the model parameters (Appendix Table~\ref{tab:detectors}).}
  \label{fig:result1}
\end{wrapfigure}

%Table~\ref{tab:in_domain} and Figure~\ref{fig:result1} shows the results of Default setting, with detailed per-model and per-generator breakdowns reported in Appendix~\ref{sec: results}. 
In the \emph{Default} setting, we evaluate detection performance on human-composed (negative) versus AI-generated (positive) music using \BenchName. Appendix Table~\ref{tab:in_domain} and Figure~\ref{fig:result1} summarize performance under the \emph{Default} setting, with detailed per-model and per-generator results provided in Appendix~\ref{sec: results}.
Overall, we observe that convolutional and transformer-based spectrogram models achieve near-perfect performance. 
Models such as ResNet18+Spec and AST attain near-perfect $\mathrm{ACC_B}$, while ConvNeXt, ViT, and SpecTTTra follow closely, all operating within a narrow margin ($99.5\%$–$99.8\%$) with negligible $\mathrm{FPR}$ and $\mathrm{FNR}$.
%ResNet18+Spec and AST are nearly perfect on $\mathrm{ACC_B}$,. ConvNeXt, ViT, and SpecTTTra follow closely, with overall balanced accuracies between $99.5\%$ and $99.8\%$, differing only by rounding error, and correspondingly negligible FPR and FNR values. 
In contrast, raw waveform CNNs, while still competitive, exhibit a modest performance gap. M5 achieves balanced accuracy of $94.3\%$ ($\mathrm{FNR}=1.4\%$), whereas RawNet2 is primarily limited by a higher $\mathrm{FPR}$ of $31.6\%$ ($\mathrm{ACC_B}=82.3\%$). 
%Deezer's handcrafted feature detector reaches $\mathrm{ACC_B}=97.6\%$, while Q-SVM performs substantially worse, with $\mathrm{ACC_B}=68.9\%$. 
A similar trend is observed for classical approaches: Deezer’s handcrafted feature-based detector performs robustly ($\mathrm{ACC_B}=97.6\%$), while Q-SVM shows substantially weaker discrimination ability ($\mathrm{ACC_B}=68.9\%$).
%Among commercial detectors, Ircam Amplify maintains high precision ($98.6\%$) but lower recall ($71.1\%$; $\mathrm{FNR}=28.9\%$), while Resemble.AI exhibits the weakest performance overall ($\mathrm{ACC_B}=46.9\%$, $\mathrm{FPR}=70.8\%$). We include commercial detectors as reference points, rather than as strictly controlled baselines against our trained detectors. 
Commercial systems exhibit mixed behavior. \emph{Ircam Amplify} maintains high precision ($98.6\%$) but suffers from lower recall ($71.1\%$, $\mathrm{FNR}=28.9\%$), indicating conservative predictions, whereas \emph{Resemble AI} performs poorly overall ($\mathrm{ACC_B}=46.9\%$, $\mathrm{FPR}=70.8\%$). These systems are included as reference points rather than strictly controlled baselines.
Our benchmark is also more challenging than prior datasets that focus primarily on vocal AI music~\cite{rahman2025sonics}, or datasets with more limited coverage of contemporary music generators~\cite{li2024m6, comanducci2024fakemusiccaps}. 
%We provide a deeper discussion of the factors influencing detection accuracy and their practical deployment implications in Section~\ref{sec: Design Insights}.
\vspace{-3mm}
\subsubsection{Generalization to Unseen Datasets}
\vspace{-3mm}
% : in each round, one generator is excluded entirely from training, and the detector is then evaluated on music produced by that unseen generator, along with human compositions, to measure both false positive and false negative rates. 
Detailed \textit{Leave-One-Model-Out (LOMO)} results are reported in Appendix Table~\ref{tab:leave_one_out} and Figure~\ref{fig:result1}. 
Overall, spectrogram-based models, including ResNet18+Spec, AST, and ConvNeXt, maintain strong performance across unseen generators, often exceeding $95\%$ balanced accuracy.
Their strong performance suggests that time--frequency representations capture generative artifacts that transfer across models. 
Raw waveform-based models show more variable behavior: M5 remains reasonably strong, whereas RawNet2 fluctuates substantially across unseen generators. 
The statistical models Q-SVM and Deezer's method, with their extracted feature representations, show unstable performance across left-out generators. 
One possible explanation for the uneven \textit{LOMO} transferability across generators is shared training-data provenance among some generator families. 
MusicGen, MAGNeT, and JASCO are all trained on licensed music catalogs from Shutterstock~\cite{shutterstock} and Pond5~\cite{pond5}, supplemented by Meta-internal collections~\cite{copet2023simple,ziv2024masked,tal2024joint}. 
This shared corpus may introduce distributional homogeneity across their outputs, helping detectors trained on one model generalize to the others. 
Similarly, Mustango~\cite{melechovsky2024mustango} and AudioLDM2~\cite{liu2024audioldm2} share lineage through AudioSet~\cite{gemmeke2017audio}: Mustango expands MusicCaps~\cite{agostinelli2023musiclm}, a subset of AudioSet, into its MusicBench corpus with richer annotations, while AudioLDM2 trains directly on AudioSet and related datasets. 
These shared origins may partially explain the non-trivial cross-model generalization observed under \textit{LOMO} and suggest that training-data overlap, in addition to model architecture, can shape detector transferability.

As shown in Appendix Table~\ref{tab:ood_accuracy}, the \textit{Out-Of-Domain (OOD)} human-only evaluation reveals substantial differences in false-alarm behavior across detectors. Among neural detectors, AST performs best ($79.8\%$), followed by M5 ($75.8\%$), which reveals that strong cross-generator AI detection does not necessarily imply reliable behavior on unfamiliar data. Deezer's handcrafted detector achieves the highest 94.4\% accuracy, demonstrating their proposed method of explicitly targeting deconvolution-induced spectral artifacts in music can achieve strong performance without relying on large-scale training data. 
At the same time, this result should be read alongside Table~\ref{tab:leave_one_out}, where Deezer shows high FNR on several left-out generators, suggesting that its strong \textit{OOD} human accuracy may partly reflect a conservative human-biased decision boundary rather than uniformly stronger robustness. The remaining detectors all fall below $60\%$ accuracy.

\vspace{-3mm}
\subsection{Robustness against Human-AI Co-Creations}
\vspace{-2mm}
\label{sec: cocreation-results}
The first column of Table~\ref{tab:Perturbed_full} reports detector performance on the Human-AI Co-Created subset. Because these tracks originate from human compositions and are only later modified by AI models using expert-inspired editing prompts, they preserve many human-like structural and performance cues while introducing partial AI-generated artifacts. ResNet18+Spec achieves zero FNR, followed by ViT($1.7\%$) and AST($4.8\%$), suggesting that strong spectrogram-based models can still capture AI-remixing artifacts even when the underlying composition originates from human music. 
By contrast, other detectors show higher FNR, indicating that handcrafted statistical features, narrow spectral fingerprints, or opaque commercial decision boundaries may miss partial AI involvement when the artifacts are localized or diluted by human source material. 
\vspace{-2mm}
\subsection{Robustness against Transformed Variants}
\vspace{-2mm}
\label{sec: transform-results}
\begin{table}[t]
  \centering

\vspace{-2mm}
\caption{FNRs on human-AI co-creations and transformed variants, showing the percentage of transformed AI-generated tracks misclassified as human (\%).}
\label{tab:Perturbed_full}
\begingroup
\setlength{\tabcolsep}{5pt}
\resizebox{0.85\columnwidth}{!}{%
\renewcommand{\arraystretch}{1.0}
\begin{NiceTabular}{c|c|ccc|ccc|cc}[colortbl-like]
\CodeBefore
    \rowcolor{gray!15}{1,2}
    \arrayrulecolor{black}
\Body
\toprule
\Block{2-1}{\textbf{Automatic}\\\textbf{Detector}}
& \Block{2-1}{\makecell{\textbf{Human-AI}\\\textbf{Co-created}}}
& \Block{1-3}{\textbf{Lossy Compression}} & & &
  \Block{1-3}{\textbf{Acoustic Post-Processing}} & & &
  \Block{1-2}{\textbf{Additive Perturbations}}\\
&
& \textbf{MP3} & \textbf{OGG} & \textbf{AAC}
& \textbf{EQ} & \textbf{Pitch/Time} & \textbf{Loudness}
& \textbf{Gaussian} & \textbf{Adversarial} \\
\midrule
Q-SVM & \cellcolor{softblue!42}36.7 & \cellcolor{softblue!46}32.1 & \cellcolor{softblue!46}33.3 & \cellcolor{softblue!56}24.8 & \cellcolor{softblue!46}32.9 & \cellcolor{softblue!48}31.9 & \cellcolor{softblue!48}31.1 & \cellcolor{softblue!42}36.4 & \cellcolor{softblue!36}41.7 \\
Deezer & \cellcolor{softblue!42}36.8 & \cellcolor{softblue!72}12.3 & \cellcolor{softblue!76}9.5 & \cellcolor{softblue!78}7.7 & \cellcolor{softblue!80}6.4 & \cellcolor{softblue!5}93.6 & \cellcolor{softblue!82}5.2 & \cellcolor{softblue!60}19.6 & \cellcolor{softblue!54}26.4 \\
M5 & \cellcolor{softblue!60}20.1 & \cellcolor{softblue!66}15.2 & \cellcolor{softblue!88}1.9 & \cellcolor{softblue!88}1.9 & \cellcolor{softblue!88}2.0 & \cellcolor{softblue!54}26.2 & \cellcolor{softblue!90}1.3 & \cellcolor{softblue!88}2.1 & \cellcolor{softblue!5}92.4 \\
RawNet2 & \cellcolor{softblue!64}16.9 & \cellcolor{softblue!80}6.0 & \cellcolor{softblue!76}8.0 & \cellcolor{softblue!82}5.1 & \cellcolor{softblue!84}4.4 & \cellcolor{softblue!82}5.3 & \cellcolor{softblue!84}3.9 & \cellcolor{softblue!84}3.7 & \cellcolor{softblue!6}88.1 \\
ConvNeXt & \cellcolor{softblue!70}12.8 & \cellcolor{softblue!90}\textbf{0.2} & \cellcolor{softblue!90}\textbf{0.3} & \cellcolor{softblue!90}0.6 & \cellcolor{softblue!90}0.2 & \cellcolor{softblue!64}17.0 & \cellcolor{softblue!90}0.2 & \cellcolor{softblue!90}0.4 & \cellcolor{softblue!72}11.7 \\
ResNet18+Spec & \cellcolor{softblue!90}\textbf{0.0} & \cellcolor{softblue!90}0.9 & \cellcolor{softblue!84}3.7 & \cellcolor{softblue!90}\textbf{0.1} & \cellcolor{softblue!86}3.1 & \cellcolor{softblue!72}11.3 & \cellcolor{softblue!88}2.0 & \cellcolor{softblue!82}5.2 & \cellcolor{softblue!72}11.1 \\
AST & \cellcolor{softblue!82}4.8 & \cellcolor{softblue!56}24.9 & \cellcolor{softblue!78}7.3 & \cellcolor{softblue!84}3.5 & \cellcolor{softblue!90}\textbf{0.0} & \cellcolor{softblue!84}\textbf{4.2} & \cellcolor{softblue!90}\textbf{0.0} & \cellcolor{softblue!90}\textbf{0.0} & \cellcolor{softblue!76}\textbf{9.2} \\
ViT & \cellcolor{softblue!88}1.7 & \cellcolor{softblue!90}0.4 & \cellcolor{softblue!90}0.5 & \cellcolor{softblue!90}0.5 & \cellcolor{softblue!90}0.2 & \cellcolor{softblue!60}19.9 & \cellcolor{softblue!90}0.2 & \cellcolor{softblue!90}0.8 & \cellcolor{softblue!74}10.7 \\
SpecTTTra & \cellcolor{softblue!46}33.6 & \cellcolor{softblue!90}0.5 & \cellcolor{softblue!90}1.0 & \cellcolor{softblue!90}1.0 & \cellcolor{softblue!90}0.5 & \cellcolor{softblue!64}16.7 & \cellcolor{softblue!90}0.5 & \cellcolor{softblue!90}0.9 & \cellcolor{softblue!74}10.7 \\
Ircam Amplify & \cellcolor{softblue!38}39.6 & \cellcolor{softblue!20}63.6 & \cellcolor{softblue!28}54.3 & \cellcolor{softblue!54}26.2 & \cellcolor{softblue!36}41.6 & \cellcolor{softblue!36}42.1 & \cellcolor{softblue!36}41.9 & \cellcolor{softblue!32}46.2 & \cellcolor{softblue!10}79.4 \\
Resemble AI & \cellcolor{softblue!60}21.0 & \cellcolor{softblue!52}27.7 & \cellcolor{softblue!42}37.4 & \cellcolor{softblue!48}30.7 & \cellcolor{softblue!36}42.0 & \cellcolor{softblue!50}29.8 & \cellcolor{softblue!50}28.8 & \cellcolor{softblue!46}32.1 & \cellcolor{softblue!38}39.3 \\
\bottomrule
\end{NiceTabular}
}
\endgroup
\vspace{-4mm}
\end{table}
\noindent\textbf{Impact of Lossy Compression.}
The second column group in Table~\ref{tab:Perturbed_full} evaluates robustness under MP3, OGG, and AAC conversions.
Most strong neural detectors remain stable under OGG and AAC, but MP3 substantially increases FNR for AST ($24.9\%$) and especially Ircam Amplify ($63.6\%$). The commercial detectors are particularly sensitive to compression, with Ircam Amplify showing high FNR under both MP3 and OGG, and Resemble AI remaining consistently weak across all three codecs. By contrast, ConvNeXt, ViT, SpecTTTra, and ResNet18+Spec remain near-zero under most compressed variants, suggesting better invariance to codec-induced artifacts.

\noindent\textbf{Impact of Acoustic Post-processing.}
EQ and loudness normalization have limited effect on the strongest spectrogram models, but pitch/time modification is highly disruptive for specific detectors. 
The most striking case is Deezer, whose FNR rises to $93.6\%$ under pitch/time modification, despite remaining low under EQ and loudness normalization. 
This is consistent with Deezer's reliance on frequency-domain fingerprints: shifting pitch or temporal structure can directly disturb the spectral periodicity it uses for detection. 
ConvNeXt, ViT, and SpecTTTra also show moderate degradation under pitch/time modification, while AST is comparatively stable, suggesting that not all spectrogram-based models rely on the same spectral cues.

\noindent\textbf{Impact of Additive Perturbations.}
Gaussian noise is relatively mild for most neural detectors, but it still increases FNR for Q-SVM, Deezer, and the commercial detectors.
The PGD attack is crafted using M5 and RawNet2 as surrogate models; their large FNR increases are expected and confirm that the attack is effective on the source models. Because the PGD perturbations are crafted against neural waveform-input surrogate models, we focus on their transfer to other neural architectures rather than to handcrafted or statistical baselines. The results show they transfer weakly to spectrogram-based neural detectors, with remaining around $9$--$12\%$ FNR.
\vspace{-3mm}
\section{Design Insights for Robust Music-Authenticity Detection}
\vspace{-3mm}
\label{sec: Design Insights}
Our results suggest that robust AI-music detection should be designed not only around stronger neural backbones, but also around realistic bypass behaviors, artifact-aware feature construction, and deployment constraints. 

\textbf{Robustness to Real-World Bypass Behaviors.}
Users attempting to evade detection are unlikely to submit only clean text-to-audio outputs. They may use human--AI co-creation, prompt-based regeneration, compression, acoustic post-processing, pitch/time modification, or additive perturbations to obscure generator traces. Our co-created music and transformed-variant experiments show that such scenarios can substantially affect detector performance. Therefore, training/evaluation datasets should explicitly include diverse transformations through artifact-aware data augmentation. 

\textbf{Handcrafted Artifact Features.} Handcrafted artifact features remain useful, especially for scalable detection. Deezer's detector~\cite{afchar2025fourier} uses periodic high-frequency spectrogram peaks as generation-related features and achieves $97.6\%$ balanced accuracy in our default experiment. Our analysis in Section~\ref{sec:sec5-artifacts} further indicates that spectrogram-level artifacts and human-interpretable acoustic cues provide meaningful detection evidence. Such features can be paired with lightweight classifiers, including logistic regression, SVMs, or compact CNNs, rather than relying on large deep neural models. 

\textbf{Benefits and Costs of Longer Audio Context.} Appendix~\ref{subsec:30s} shows that longer audio segments improve detection performance, likely because they expose temporal structure, arrangement consistency, vocal transitions, and repetition patterns that may be absent from short clips. However, longer inputs also increase inference cost. Practical detectors therefore need to balance the stronger evidence provided by longer excerpts against latency and computational constraints. 

\textbf{Design for Scalable Deployment.}
Real-world platforms such as Spotify, YouTube, SoundCloud, and TikTok ingest massive volumes of audio, making it expensive and latency-sensitive to run heavy detectors on every upload. As shown in Appendix~\ref{subsec:inference_time}, detectors vary substantially in per-sample inference time, and Appendix Table~\ref{tab:detectors} further highlights their architectural differences. Lightweight detectors can serve as first-stage filters that screen every track in real time or near real time, reduce cloud-computing cost, and scale to edge servers, content delivery network nodes, or on-device environments where compute is limited ~\cite{ahmed2020void}.

\vspace{-4mm}
\section{Identifying Artifacts in AI-Generated Music}
\label{sec:sec5-artifacts}
\vspace{-3mm}
Beyond evaluating machine detector capability, we also examine signal-level, spectral, structural, vocal, and lyric-related cues that reveal synthesis irregularities and help explain how AI-generated music diverges from human compositions.

\textbf{Frequency-Sensitivity Analysis}
To understand which frequency regions most influence AI-music detection, we selectively suppress narrow spectral bands and observing how detector performance changes. Specifically, a 4th-order Butterworth band-stop filter is applied to one sub-band at a time, and each model's output is compared with its full-band baseline across four sets of AI-generated music in \BenchName, using seven detectors. Results for SpecTTTra and ResNet18+Spec are shown in Appendix Figure~\ref{fig:bandstop}, with others in Appendix~\ref{subsubsec:freq}. To visualize how detectors attend to different spectral regions, we apply Grad-CAM analysis~\cite{selvaraju2017grad} on ResNet18+Spec using representative tracks from both \textit{vocal} and \textit{instrumental} sets (Appendix Figure~\ref{fig:gradcam}). This analysis provides a diagnostic view of the frequency bands each detector relies on when making predictions. By identifying the bands whose suppression most changes detector outputs, the spectral regions carrying the strongest detection evidence can be localized, enabling subsequent artifact analysis to focus on those regions.

\textbf{Deconvolution Artifacts and Second-Order Spectral Analysis}
Artifacts from transposed convolutions, a.k.a., deconvolution or checkerboard artifacts, were first observed in image synthesis~\cite{odena2016checkerboard} and later found to produce analogous spectral irregularities in audio~\cite{afchar2025fourier, donahue2019wavegan,pons2021upsampling}. These arise when the kernel size and stride interact unevenly, creating periodic amplitude modulations that manifest as subtle spectral line patterns. Many music generation models, including those using the HiFi-GAN~\cite{kong2020hifigan} or neural compressors such as EnCodec~\cite{defossez2022encodec} (e.g., MusicGen, MAGNeT), inherently perform such transposed convolution upsampling, leaving similar structural fingerprints in their outputs. To further analyze this effect, we probed EnCodec with a 1s unit-impulse (details in Appendix~\ref{sec:encodec}), whose reconstruction reveals clear periodic amplitude modulations. Such periodic modulations are often masked by rich musical content and remain difficult to isolate using standard magnitude spectrum. In our analysis, we find that applying a second-order Fast Fourier Transform (FFT), i.e., the FFT of the magnitude spectrum, reveals these artifacts far more distinctly, as shown in Appendix Figure~\ref{fig:secondfft}.

\textbf{Chord Progression Artifacts}
Another artifact arises from the mismatch between locally plausible audio signals and globally coherent musical structure. 
Text-to-music generators are typically optimized in waveform, spectrogram, or latent audio spaces, and may not explicitly model harmonic theory, genre conventions, or instrument-specific performance practice. 
Although music has no universal chord standard, trained musicians can use diagnostic cues for identifying stylistically implausible chord progressions\footnote{https://www.guitartricks.com/blog/Rock-Guitar-Chord-Systems.}\footnote{https://blog.native-instruments.com/rock-chord-progressions/.}. 
We therefore use chord-level analysis as a human-interpretable diagnostic by extracting chord sequences from AI-generated rock tracks with \texttt{chord-extractor}\footnote{https://github.com/ohollo/chord-extractor.}. 
Representative examples in Appendix~\ref{sec:chord} show abrupt or stylistically inconsistent chord transitions.

\textbf{Performance-Constraint Artifacts}
We also identify artifacts that violate the physical limits of human performance and instrument design. AI-generated pieces occasionally include events that are mechanically impossible, such as piano passages exceeding ten simultaneous notes, guitar phrases extending beyond the instrument's fretboard range, or wind lines that disregard breathing constraints, with representative examples provided in Appendix~\ref{sec:performance-constraint}.

\textbf{Lyrical Artifacts} AI-generated singing often exhibits linguistic artifacts that undermine intelligibility and vocal realism. We identify two main types: \textit{pseudo-lyric artifacts} and \textit{approximate-pronunciation artifacts}. Pseudo-lyric artifacts occur when models generate syllable-like sequences that match musical rhythm but lack semantic meaning; approximate-pronunciation artifacts appear when lyrics are partially reproduced but acoustically distorted, with representative examples in Appendix~\ref{sec:lyrics}.

Taken together, these analyses show that artifact inspection can provide actionable signals for detector design. Future detectors can incorporate frequency-band sensitivity, second-order spectral patterns, chord-transition irregularities, physical-performance violations, and lyrical artifacts as explicit feature channels or auxiliary supervision signals. Once extracted, these artifact cues can be represented as structured feature vectors and used to train lightweight classifiers or regression models, providing an interpretable alternative or complement to end-to-end neural detectors.
\vspace{-3mm}
\section{Conclusion}
\vspace{-3mm}

This work introduced a unified framework for evaluating how machines detect and interpret AI-generated music.
By combining large-scale detector benchmarking with systematic artifact analysis, we reveal key differences in how automated systems identify AI-generated music across clean, transformed, and out-of-domain settings.
The framework, dataset, and analyses together provide a reproducible foundation for advancing robust and interpretable detection systems.
Importantly, the artifacts uncovered in our acoustic analyses provide valuable guidance for designing next-generation AI-music detectors that align more closely with musical structure.

\bibliographystyle{plain} 
\bibliography{references} 

@String{Computing = "Computing" }

@String{Computer = "{IEEE} Computer" }

@misc{geniuslyrics,
  author       = {CarlosGDCJ},
  title        = {Genius song lyrics with language information},
  howpublished = {\url{https://www.kaggle.com/datasets/carlosgdcj/genius-song-lyrics-with-language-information}},
  year         = {2023},
  note         = {Accessed: 2025-09-29}
}

@article{agarwal2025gpt,
  title={{Gpt-oss-120b \& gpt-oss-20b model card}},
  author={Agarwal, Sandhini and Ahmad, Lama and Ai, Jason and Altman, Sam and Applebaum, Andy and Arbus, Edwin and Arora, Rahul K and Bai, Yu and Baker, Bowen and Bao, Haiming and others},
  journal={arXiv preprint arXiv:2508.10925},
  year={2025}
}

@article{roy2025jamendomaxcaps,
  title={{JamendoMaxCaps: A Large Scale Music-caption Dataset with Imputed Metadata}},
  author={Roy, Abhinaba and Liu, Renhang and Lu, Tongyu and Herremans, Dorien},
  journal={arXiv:2502.07461},
  year={2025}
}

@ArtifactSoftware{R,
    title = {R: A Language and Environment for Statistical Computing},
    author = {{R Core Team}},
    organization = {R Foundation for Statistical Computing},
    address = {Vienna, Austria},
    year = {2019},
    url = {https://www.R-project.org/},
}

@article{agostinelli2023musiclm,
  title={{MusicLM: Generating Music from Text}},
  author={Agostinelli, Andrea and Denk, Timo I and Borsos, Zal{\'a}n and Engel, Jesse and Verzetti, Mauro and Caillon, Antoine and Huang, Qingqing and Jansen, Aren and Roberts, Adam and Tagliasacchi, Marco and others},
  journal={arXiv preprint arXiv:2301.11325},
  year={2023}
}

@article{ziv2024masked,
  title={{Masked audio generation using a single non-autoregressive transformer}},
  author={Ziv, Alon and Gat, Itai and Lan, Gael Le and Remez, Tal and Kreuk, Felix and D{\'e}fossez, Alexandre and Copet, Jade and Synnaeve, Gabriel and Adi, Yossi},
  journal={arXiv preprint arXiv:2401.04577},
  year={2024}
}

@misc{Forsgren_Martiros_2022,
  author       = {Seth Forsgren and Hayk Martiros},
  title        = {{Riffusion: Stable Diffusion for Real-time Music Generation}},
  howpublished = {\url{https://riffusion.com/about}},
  year         = {2022}
}

@misc{lyriateam2025livemusicmodels,
      title={{Live Music Models}}, 
      author={Lyria Team and Antoine Caillon and Brian McWilliams and Cassie Tarakajian and Ian Simon and Ilaria Manco and Jesse Engel and Noah Constant and Yunpeng Li and Timo I. Denk and Alberto Lalama and Andrea Agostinelli and Cheng-Zhi Anna Huang and Ethan Manilow and George Brower and Hakan Erdogan and Heidi Lei and Itai Rolnick and Ivan Grishchenko and Manu Orsini and Matej Kastelic and Mauricio Zuluaga and Mauro Verzetti and Michael Dooley and Ondrej Skopek and Rafael Ferrer and Zalán Borsos and Äaron van den Oord and Douglas Eck and Eli Collins and Jason Baldridge and Tom Hume and Chris Donahue and Kehang Han and Adam Roberts},
      year={2025},
      eprint={2508.04651},
      archivePrefix={arXiv},
      primaryClass={cs.SD},
      url={https://arxiv.org/abs/2508.04651}, 
}

@article{tal2024joint,
  title={{Joint audio and symbolic conditioning for temporally controlled text-to-music generation}},
  author={Tal, Or and Ziv, Alon and Gat, Itai and Kreuk, Felix and Adi, Yossi},
  journal={arXiv preprint arXiv:2406.10970},
  year={2024}
}

@article{madry2017towards,
  title={{Towards deep learning models resistant to adversarial attacks}},
  author={Madry, Aleksander and Makelov, Aleksandar and Schmidt, Ludwig and Tsipras, Dimitris and Vladu, Adrian},
  journal={arXiv preprint arXiv:1706.06083},
  year={2017}
}

@article{copet2023simple,
  title={{Simple and Controllable Music Generation}},
  author={Copet, Jade and Kreuk, Felix and Gat, Itai and Remez, Tal and Kant, David and Synnaeve, Gabriel and Adi, Yossi and D{\'e}fossez, Alexandre},
  journal={Advances in Neural Information Processing Systems},
  volume={36},
  pages={47704--47720},
  year={2023}
}

@misc{sunoai2023,
  author = {{Suno, Inc.}},
  title = {{Suno: AI Music Generator}},
  howpublished = {\url{https://suno.com}},
  month={[Accessed: 2025-11-13]}
}

@misc{udio2024,
  author = {{Udio, Inc.}},
  title = {{Udio: AI Music Generator}},
  howpublished = {\url{https://www.udio.com}},
  month={[Accessed: 2025-11-13]}
}

@article{doh2023lp,
  title={{Lp-musiccaps: Llm-based pseudo music captioning}},
  author={Doh, SeungHeon and Choi, Keunwoo and Lee, Jongpil and Nam, Juhan},
  journal={arXiv preprint arXiv:2307.16372},
  year={2023}
}

@inproceedings{evans2025stable,
  title={{Stable Audio Open}},
  author={Evans, Zach and Parker, Julian D and Carr, CJ and Zukowski, Zack and Taylor, Josiah and Pons, Jordi},
  booktitle={ICASSP 2025-2025 IEEE International Conference on Acoustics, Speech and Signal Processing (ICASSP)},
  pages={1--5},
  year={2025},
  organization={IEEE}
}

@misc{nytimesAIFake,
  author       = {Joe Coscarelli},
  title        = {{An A.I. Hit of Fake ‘Drake’ and ‘The Weeknd’ Rattles the Music World}},
  howpublished = {\url{https://www.nytimes.com/2023/04/19/arts/music/ai-drake-the-weeknd-fake.html}},
  note         = {Accessed: 2025-11-12},
  year         = {2023}
}

@book{gervais2015collective,
  title={{Collective management of copyright and related rights}},
  author={Gervais, Daniel},
  year={2015},
  publisher={Kluwer Law International BV}
}

@misc{recordingacademyArtificialIntelligence,
  author       = {{The Recording Academy}},
  title        = {{Artificial Intelligence Hub}},
  howpublished = {\url{https://www.recordingacademy.com/about/artificial-intelligence}},
  month        = {[Accessed: 2025-04-06]},
  year         = {2023}
}

@misc{musicbusinessworldwideItsTime,
  author       = {{Music Business Worldwide}},
  title        = {{It's Time for Streaming Services to Act on AI Music}},
  howpublished = {\url{https://www.musicbusinessworldwide.com/its-time-for-streaming-services-to-act-on-ai-music}},
  note         = {Accessed: 2025-11-06},
  year         = {2024}
}

@article{comanducci2024fakemusiccaps,
  title={{FakeMusicCaps: a Dataset for Detection and Attribution of Synthetic Music Generated via Text-to-Music Models}},
  author={Comanducci, Luca and Bestagini, Paolo and Tubaro, Stefano},
  journal={arXiv preprint arXiv:2409.10684},
  year={2024},
  eprint={2409.10684},
  archivePrefix={arXiv},
  primaryClass={eess.AS},
  url={https://arxiv.org/abs/2409.10684}
}

@inproceedings{rahman2025sonics,
  author    = {Md Awsafur Rahman and Zaber Ibn Abdul Hakim and Najibul Haque Sarker and Bishmoy Paul and Shaikh Anowarul Fattah},
  title     = {{{SONICS}: Synthetic or Not---Identifying Counterfeit Songs}},
  booktitle = {Proceedings of the International Conference on Learning Representations (ICLR)},
  year      = {2025},
  url       = {https://arxiv.org/abs/2408.14080}
}

@inproceedings{liu2022convnet,
  title={{A convnet for the 2020s}},
  author={Liu, Zhuang and Mao, Hanzi and Wu, Chao-Yuan and Feichtenhofer, Christoph and Darrell, Trevor and Xie, Saining},
  booktitle={Proceedings of the IEEE/CVF conference on computer vision and pattern recognition},
  pages={11976--11986},
  year={2022}
}

@article{li2024m6,
  title={{M6: Multi-generator, Multi-domain, Multi-lingual and cultural, Multi-genres, Multi-instrument Machine-Generated Music Detection Databases}},
  author={Li, Yupei and Li, Hanqian and Specia, Lucia and Schuller, Bj{\"o}rn W},
  journal={arXiv preprint arXiv:2412.06001},
  year={2024},
  eprint={2412.06001},
  archivePrefix={arXiv},
  primaryClass={cs.SD},
  url={https://arxiv.org/abs/2412.06001}
}

@article{li2024xai,
  title={{Detecting Machine-Generated Music with Explainability -- A Challenge and Early Benchmarks}},
  author={Li, Yupei and Sun, Qiyang and Li, Hanqian and Specia, Lucia and Schuller, Bj{\"o}rn W},
  journal={arXiv preprint arXiv:2412.13421},
  year={2024},
  eprint={2412.13421},
  archivePrefix={arXiv},
  primaryClass={cs.SD},
  url={https://arxiv.org/abs/2412.13421}
}

@article{afchar2024detecting,
  title={{Detecting music deepfakes is easy but actually hard}},
  author={Afchar, Darius and Meseguer-Brocal, Gabriel and Hennequin, Romain},
  journal={arXiv preprint arXiv:2405.04181},
  year={2024}
}

@inproceedings{afchar2025challenges,
  title={{AI-Generated Music Detection and its Challenges}},
  author={Afchar, Darius and Meseguer-Brocal, Gabriel and Hennequin, Romain},
  booktitle={ICASSP 2025-2025 IEEE International Conference on Acoustics, Speech and Signal Processing (ICASSP)},
  pages={1--5},
  year={2025},
  organization={IEEE}
}

@article{gong2021ast,
  title={{Ast: Audio spectrogram transformer}},
  author={Gong, Yuan and Chung, Yu-An and Glass, James},
  journal={arXiv preprint arXiv:2104.01778},
  year={2021}
}

@inproceedings{ha2024organic,
  title={{Organic or diffused: Can we distinguish human art from ai-generated images?}},
  author={Ha, Anna Yoo Jeong and Passananti, Josephine and Bhaskar, Ronik and Shan, Shawn and Southen, Reid and Zheng, Haitao and Zhao, Ben Y},
  booktitle={Proceedings of the 2024 on ACM SIGSAC Conference on Computer and Communications Security},
  pages={4822--4836},
  year={2024}
}

@inproceedings{dai2017very,
  title={{Very deep convolutional neural networks for raw waveforms}},
  author={Dai, Wei and Dai, Chia and Qu, Shuhui and Li, Juncheng and Das, Samarjit},
  booktitle={IEEE International Conference on Acoustics, Speech and Signal Processing (ICASSP)},
  pages={421--425},
  year={2017},
  organization={IEEE}
}

@inproceedings{tak2021rawnet2,
  title={{End-to-end anti-spoofing with rawnet2}},
  author={Tak, Hemlata and Patino, Jose and Todisco, Massimiliano and Nautsch, Andreas and Evans, Nicholas and Larcher, Anthony},
  booktitle={ICASSP 2021-2021 IEEE International Conference on Acoustics, Speech and Signal Processing (ICASSP)},
  pages={6369--6373},
  year={2021},
  organization={IEEE}
}

@inproceedings{he2016deep,
  title={{Deep residual learning for image recognition}},
  author={He, Kaiming and Zhang, Xiangyu and Ren, Shaoqing and Sun, Jian},
  booktitle={Proceedings of the IEEE conference on computer vision and pattern recognition},
  pages={770--778},
  year={2016}
}

@inproceedings{singh2021qsvm,
  title={{Detection of AI-synthesized speech using cepstral \& bispectral statistics}},
  author={Singh, Arun Kumar and Singh, Priyanka},
  booktitle={2021 IEEE 4th International Conference on Multimedia Information Processing and Retrieval (MIPR)},
  pages={412--417},
  year={2021},
  organization={IEEE}
}

@article{dosovitskiy2021vit,
  title={{An image is worth 16x16 words: Transformers for image recognition at scale}},
  author={Dosovitskiy, Alexey and Beyer, Lucas and Kolesnikov, Alexander and Weissenborn, Dirk and Zhai, Xiaohua and Unterthiner, Thomas and Dehghani, Mostafa and Minderer, Matthias and Heigold, Georg and Gelly, Sylvain and others},
  journal={arXiv preprint arXiv:2010.11929},
  year={2021}
}

@article{liu2024audioldm2,
  title={{Audioldm 2: Learning holistic audio generation with self-supervised pretraining}},
  author={Liu, Haohe and Yuan, Yi and Liu, Xubo and Mei, Xinhao and Kong, Qiuqiang and Tian, Qiao and Wang, Yuping and Wang, Wenwu and Wang, Yuxuan and Plumbley, Mark D},
  journal={IEEE/ACM Transactions on Audio, Speech, and Language Processing},
  volume={32},
  pages={2871--2883},
  year={2024},
  publisher={IEEE}
}

@inproceedings{melechovsky2024mustango,
  author    = {Melechovsky, J. and Guo, Z. and Ghosal, D. and Majumder, N. and Herremans, D. and Poria, S.},
  title     = {{{Mustango}: Toward Controllable Text-to-Music Generation}},
  booktitle = {Proceedings of the 2024 Conference of the North American Chapter of the Association for Computational Linguistics ({NAACL})},
  year      = {2024},
  pages     = {8293--8316},
  publisher = {Association for Computational Linguistics},
}

@article{defossez2022encodec,
  title   = {{High Fidelity Neural Audio Compression}},
  author  = {D{\'e}fossez, Alexandre and Copet, Jade and Synnaeve, Gabriel and Adi, Yossi},
  journal = {arXiv preprint arXiv:2210.13438},
  year    = {2022}
}

@article{kong2020hifigan,
  title={Hifi-gan: Generative adversarial networks for efficient and high fidelity speech synthesis},
  author={Kong, Jungil and Kim, Jaehyeon and Bae, Jaekyoung},
  journal={Advances in neural information processing systems},
  volume={33},
  pages={17022--17033},
  year={2020}
}

@inproceedings{pons2021upsampling,
  title={{Upsampling artifacts in neural audio synthesis}},
  author={Pons, Jordi and Pascual, Santiago and Cengarle, Giulio and Serr{\`a}, Joan},
  booktitle={ICASSP 2021-2021 IEEE International Conference on Acoustics, Speech and Signal Processing (ICASSP)},
  pages={3005--3009},
  year={2021},
  organization={IEEE}
}

@article{odena2016checkerboard,
  title={{Deconvolution and checkerboard artifacts}},
  author={Odena, Augustus and Dumoulin, Vincent and Olah, Chris},
  journal={Distill},
  volume={1},
  number={10},
  pages={e3},
  year={2016}
}

@article{donahue2019wavegan,
  author    = {Chris Donahue and Julian McAuley and Miller Puckette},
  title     = {{Adversarial Audio Synthesis}},
  journal   = {arXiv preprint arXiv:1802.04208},
  year      = {2019},
  url       = {https://arxiv.org/abs/1802.04208}
}

@article{afchar2025fourier,
  title={{A Fourier Explanation of AI-music Artifacts}},
  author={Afchar, Darius and Meseguer-Brocal, Gabriel and Akesbi, Kamil and Hennequin, Romain},
  journal={arXiv preprint arXiv:2506.19108},
  year={2025}
}

@inproceedings{roberts2018hierarchical,
  title={{A hierarchical latent vector model for learning long-term structure in music}},
  author={Roberts, Adam and Engel, Jesse and Raffel, Colin and Hawthorne, Curtis and Eck, Douglas},
  booktitle={International conference on machine learning},
  pages={4364--4373},
  year={2018},
  organization={PMLR}
}

@article{dhariwal2020jukebox,
  title={{Jukebox: A generative model for music}},
  author={Dhariwal, Prafulla and Jun, Heewoo and Payne, Christine and Kim, Jong Wook and Radford, Alec and Sutskever, Ilya},
  journal={arXiv preprint arXiv:2005.00341},
  year={2020}
}

@inproceedings{dong2018musegan,
  title={{Musegan: Multi-track sequential generative adversarial networks for symbolic music generation and accompaniment}},
  author={Dong, Hao-Wen and Hsiao, Wen-Yi and Yang, Li-Chia and Yang, Yi-Hsuan},
  booktitle={Proceedings of the AAAI Conference on Artificial Intelligence},
  volume={32},
  number={1},
  year={2018}
}

@article{rouard2024audio,
  title={{Audio conditioning for music generation via discrete bottleneck features}},
  author={Rouard, Simon and Adi, Yossi and Copet, Jade and Roebel, Axel and D{\'e}fossez, Alexandre},
  journal={arXiv preprint arXiv:2407.12563},
  year={2024}
}

@inproceedings{cramer2019look,
  title={{Look, listen, and learn more: Design choices for deep audio embeddings}},
  author={Cramer, Aurora Linh and Wu, Ho-Hsiang and Salamon, Justin and Bello, Juan Pablo},
  booktitle={ICASSP 2019-2019 IEEE International Conference on Acoustics, Speech and Signal Processing (ICASSP)},
  pages={3852--3856},
  year={2019},
  organization={IEEE}
}

@article{frohmann2025ai,
  title={{AI-Generated Song Detection via Lyrics Transcripts}},
  author={Frohmann, Markus and Epure, Elena V and Meseguer-Brocal, Gabriel and Schedl, Markus and Hennequin, Romain},
  journal={arXiv preprint arXiv:2506.18488},
  year={2025}
}

@misc{newsroomdeezerDeezerDeploys,
  author       = {{Deezer Newsroom}},
  title        = {{Deezer deploys cutting-edge AI detection tool for music streaming}},
  howpublished = {\url{https://newsroom-deezer.com/2025/01/deezer-deploys-cutting-edge-ai-detection-tool-for-music-streaming/}},
  note         = {Accessed: 2025-08-16},
}

@misc{audiblemagicIdentifyingCover,
  author       = {{Audible Magic}},
  title        = {{Identifying cover songs, live performances, AI clones and more}},
  howpublished = {\url{https://www.audiblemagic.com/2024/02/07/identifying-cover-songs-live-performances-ai-clones-and-more/}},
  note         = {Accessed: 2025-08-16},
}

@misc{musicbusinessworldwideBelieveDeveloped,
  author  = {{Music Business Worldwide}},
  title   = {Believe has developed software that detects AI-made music with 98\% accuracy: What might this mean for the future?},
  url     = {https://www.musicbusinessworldwide.com/believe-has-developed-its-own-ai-made-music-detector-with-98-accuracy-what-might-this-mean-for-the-future/},
  urldate = {2025-08-16},
}

@misc{ircamamplifyMusicDetector,
  author       = {{Ircam Amplify}},
  title        = {{AI Music Detector}},
  howpublished = {\url{https://www.ircamamplify.io/product/ai-generated-music-detector}},
  note         = {Accessed: 2025-08-19},
}

@misc{resembleDetectDeepfakes,
  author       = {{Resemble AI}},
  title        = {{Detect Deepfakes with Resemble}},
  howpublished = {\url{https://www.resemble.ai/detect/}},
  note         = {Accessed: 2025-08-19},
}

@article{titze2021effect,
  title={{The effect of single harmonic tuning on vocal loudness}},
  author={Titze, Ingo R},
  journal={Journal of Voice},
  volume={35},
  number={6},
  pages={832--837},
  year={2021},
  publisher={Elsevier}
}

@article{Pont2019CAA_SibilantS,
  title={{Computational aeroacoustics to identify sound sources in the generation of sibilant/s}},
  author={Pont, Arnau and Guasch, Oriol and Baiges, Joan and Codina, Ramon and van Hirtum, Annemie},
  journal={International journal for numerical methods in biomedical engineering},
  volume={35},
  number={1},
  pages={e3153},
  year={2019},
  publisher={Wiley Online Library}
}

@inproceedings{LeRoux2008SAPA,
  title={{Explicit consistency constraints for STFT spectrograms and their application to phase reconstruction}},
  author={Le Roux, Jonathan and Ono, Nobutaka and Sagayama, Shigeki},
  booktitle={SAPA@ INTERSPEECH},
  pages={23--28},
  year={2008}
}

@article{Zhu2024MusicHiFi,
  title={MusicHiFi: Fast high-fidelity stereo vocoding},
  author={Zhu, Ge and Caceres, Juan-Pablo and Duan, Zhiyao and Bryan, Nicholas J},
  journal={IEEE Signal Processing Letters},
  year={2024},
  publisher={IEEE}
}

@article{Marafioti2019ContextEncoder,
  title={{A context encoder for audio inpainting}},
  author={Marafioti, Andr{\'e}s and Perraudin, Nathana{\"e}l and Holighaus, Nicki and Majdak, Piotr},
  journal={IEEE/ACM Transactions on Audio, Speech, and Language Processing},
  volume={27},
  number={12},
  pages={2362--2372},
  year={2019},
  publisher={IEEE}
}

@misc{WikipediaWhistleRegister,
  author       = {{Wikipedia contributors}},
  title        = {{Whistle register --- {Wikipedia}{,} The Free Encyclopedia}},
  year         = {2025},
  url          = {https://en.wikipedia.org/wiki/Whistle_register},
  note         = {Accessed: 2025-11-07}
}

@inproceedings{selvaraju2017grad,
  title={{Grad-cam: Visual explanations from deep networks via gradient-based localization}},
  author={Selvaraju, Ramprasaath R and Cogswell, Michael and Das, Abhishek and Vedantam, Ramakrishna and Parikh, Devi and Batra, Dhruv},
  booktitle={Proceedings of the IEEE international conference on computer vision},
  pages={618--626},
  year={2017}
}

@article{brezas2024vibrational,
  title={{Vibrational Analysis of a Splash Cymbal by Experimental Measurements and Parametric CAD-FEM Simulations}},
  author={Brezas, Spyros and Kaselouris, Evaggelos and Orphanos, Yannis and Tatarakis, Michael and Bakarezos, Makis and Papadogiannis, Nektarios A and Dimitriou, Vasilis},
  journal={Vibration},
  volume={7},
  number={1},
  pages={146--160},
  year={2024},
  publisher={MDPI}
}

@inproceedings{Petrosino2016,
  title={{Ultrasonic components of musical instruments}},
  author={Petrosino, Jorge and Canalis, Ianina},
  booktitle={Proceedings of Meetings on Acoustics},
  volume={28},
  number={1},
  pages={035006},
  year={2016},
  organization={Acoustical Society of America}
}

@misc{UNSW_Harmonics,
  author       = {Joe Wolfe},
  title        = {{How harmonic are harmonics?}},
  howpublished = {\url{https://www.phys.unsw.edu.au/jw/harmonics.html}},
  month        = {[Accessed: 2025-11-10]}
}

@inproceedings{gemmeke2017audio,
  title={{Audio set: An ontology and human-labeled dataset for audio events}},
  author={Gemmeke, Jort F and Ellis, Daniel PW and Freedman, Dylan and Jansen, Aren and Lawrence, Wade and Moore, R Channing and Plakal, Manoj and Ritter, Marvin},
  booktitle={2017 IEEE international conference on acoustics, speech and signal processing (ICASSP)},
  pages={776--780},
  year={2017},
  organization={IEEE}
}

@misc{shutterstock, author = {Shutterstock}, 
title = {{Exclusive Royalty Free Stock Music}}, 
howpublished = {\url{https://www.shutterstock.com/music}}, 
note = {Accessed: 2025-11-12}}

@misc{pond5, author = {Pond5}, 
title = {{Download 4K and HD royalty-free stock video}}, 
howpublished = {\url{https://www.pond5.com/}}, note = {Accessed: 2025-11-12} }

@article{sunoRevenue,
  title   = {Suno: We've hit 2M paid subscribers and \$300M annual revenue},
  author  = {Stassen, Murray},
  journal = {Music Business Worldwide},
  year    = {2026},
  month   = feb,
  url     = {https://www.musicbusinessworldwide.com/suno-hits-2m-paid-subscribers-300m-annual-revenue/}
}

@misc{spotifyRoyalties,
  title        = {Royalties Guide},
  author       = {{Spotify for Artists}},
  year         = {2026},
  howpublished = {\url{https://artists.spotify.com/en/royalties-guide}},
  note         = {Accessed: 2026-05-01}
}

@article{defferrard2016fma,
  title={FMA: A dataset for music analysis},
  author={Defferrard, Micha{\"e}l and Benzi, Kirell and Vandergheynst, Pierre and Bresson, Xavier},
  journal={arXiv preprint arXiv:1612.01840},
  year={2016}
}

@inproceedings{ahmed2020void,
  title={Void: A fast and light voice liveness detection system},
  author={Ahmed, Muhammad Ejaz and Kwak, Il-Youp and Huh, Jun Ho and Kim, Iljoo and Oh, Taekkyung and Kim, Hyoungshick},
  booktitle={29th USENIX Security Symposium (USENIX Security 20)},
  pages={2685--2702},
  year={2020}
}

@misc{resemble_music, author = {Resemble.AI}, 
title = {{How to Detect AI-Generated Music Using Tools}}, 
howpublished = {\url{https://www.resemble.ai/resources/ai-music-detector-tools}}, 
note = {Accessed: 2026-05-06}}
%%%%%%%%%%%%%%%%%%%%%%%%%%%%%%%%%%%%%%%%%%%%%%%%%%%%%%%%%%%%
\clearpage

\appendix
\section*{Appendix Table of Contents}
\startcontents[sections]
\printcontents[sections]{l}{1}{\setcounter{tocdepth}{2}}

%%%%%%%%%%%%%%%%%%%%%%%%%%%%%%%%%%%%%%%%%%%%%%%%%%%%%%%%%%%%
\clearpage

\clearpage
\section{Music generation pipeline}
\label{sec: generation-pipeline}

\begin{table}[t]
  \centering
  \caption{Benchmark dataset comparison.}
  \label{tab:dataset_comparison}
  \normalsize

  % ---- table styling ----
  \rowcolors{2}{white}{lightblue}
  \renewcommand{\arraystretch}{1.4}
  \setlength{\tabcolsep}{6pt}

  \resizebox{0.98\linewidth}{!}{%
  \begin{tabular}{
      >{\centering\arraybackslash}m{4cm}  % Row labels (attributes)
      >{\centering\arraybackslash}m{2.2cm}    % M6
      >{\centering\arraybackslash}m{3cm}    % FakeMusicCaps
      >{\centering\arraybackslash}m{2.2cm}    % SONICS
      >{\centering\arraybackslash}m{2.2cm}    % BenchName
  }

    % ================= Header row =================
    \rowcolor{darkblue}
    {\color{white}\textbf{Dimension}} &
    {\color{white}\textbf{M6~\cite{li2024m6}}} &
    {\color{white}\textbf{FakeMusicCaps~\cite{comanducci2024fakemusiccaps}}} &
    {\color{white}\textbf{SONICS~\cite{rahman2025sonics}}} &
    {\color{white}\textbf{\BenchName}} \\

    \textbf{\# Generators} & 3 & 5 & 2 & \textbf{10} \\
    \textbf{Musical Coverage} & Vocal/Instr. & Vocal/Instr. & Vocal & Vocal/Instr. \\
    \textbf{Human-AI Co-Creations} & \xmark & \xmark & \xmark & \cmark \\
    \textbf{Lossy Compression} & \xmark & \xmark & \xmark & \cmark \\
    \textbf{Acoustic Post-Processing} & \xmark & \xmark & \xmark & \cmark \\
    \textbf{Additive Perturbation} & \xmark & \xmark & \xmark & \cmark \\
    \textbf{Total Tracks} & 13{,}493 & 33{,}126 & 97{,}164 & \textbf{361,166} \\

    \bottomrule
  \end{tabular}%
  }
\end{table}

\begin{table}[t]
  \centering
  \caption{Specifications for each text-to-music generator.}
  \label{tab:generatorsversion}
  \large % enlarged to normal text size

  % ---- table styling ----
  \rowcolors{2}{white}{lightblue} % alternating row colors
  \renewcommand{\arraystretch}{1.4} % increase row height for better readability
  \setlength{\tabcolsep}{6pt} % widen column spacing

  \resizebox{0.85\linewidth}{!}{%
  \begin{tabular}{
      >{\centering\arraybackslash}m{4cm}  % Generator
      >{\centering\arraybackslash}m{2cm}  % Year
      >{\centering\arraybackslash}m{6cm}  % Version
      >{\centering\arraybackslash}m{3cm}  % Backbone
      >{\centering\arraybackslash}m{1.8cm}  % Params
  }
    % ===== Header row =====
    \rowcolor{darkblue}
    {\color{white}\textbf{Generator}} &
    {\color{white}\textbf{Year}} &
    {\color{white}\textbf{Version / Checkpoint}} &
    {\color{white}\textbf{Backbone}} &
    {\color{white}\textbf{\#Params}} \\
    MusicGen~\cite{copet2023simple} & 2023 & musicgen-medium-1.5B & Transformer & 1.5B \\
    AudioLDM2~\cite{liu2024audioldm2} & 2024 & audioldm2-music-665k & Diffusion & 665M \\
    Riffusion~\cite{Forsgren_Martiros_2022} & 2022 & riffusion-model-v1 & Diffusion & 860M \\
    MAGNeT~\cite{ziv2024masked} & 2024 & magnet-medium-30secs & Transformer & 1.5B \\
    JASCO~\cite{tal2024joint} & 2024 & jasco-chords-drums-melody-400M & Flow-Matching & 400M \\
    Stable Audio Open~\cite{evans2025stable} & 2025 & stable-audio-open-1.0 & Diffusion & 1.1B \\
    Mustango~\cite{melechovsky2024mustango} & 2024 & mustango-pretrained & Diffusion & 1.4B \\
    Lyria Realtime~\cite{lyriateam2025livemusicmodels} & Up-to-date & API & Transformer & --- \\
    Suno~\cite{sunoai2023} & Up-to-date & Suno v2, v3, v3.5 & Proprietary & --- \\
    Udio~\cite{udio2024} & Up-to-date & Udio 32, 130 & Proprietary & --- \\
    \bottomrule
  \end{tabular}%
  }
\end{table}
\subsection{Challenges and Design Dimensions}
\label{subsec:objectives_challenges}

We present an early comprehensive analysis of AI-music detection, designed to capture various \textit{interacting dimensions} under clean provenance and standardized, reproducible evaluation. 
The dimensions we consider include:

\begin{itemize}[leftmargin=*, topsep=0pt, itemsep=0pt, parsep=0pt]
\item \textbf{Musical Coverage:} A comprehensive set of human, AI-generated, and AI–human co-created tracks across diverse genres and styles, including vocal and instrumental forms, to minimize genre and production bias.

\item \textbf{Generator Breadth:} A diverse set of state-of-the-art generative models spanning different architectures to enable comprehensive and architecture-agnostic evaluation of detection methods.

\item \textbf{Style Alignment:} A detailed set of captions extracted from human music serves as prompts for AI music generation to ensure that the evaluation focuses on generative characteristics rather than stylistic differences.

\item \textbf{Detector Diversity:} A portfolio of detectors across architectural families to assess detection accuracy, robustness, and cross-generator generalization, etc.

\item \textbf{Transformations \& Attacks:} Real-world audio transformations and adversarial perturbations that can obscure detection cues, stress-testing detector robustness.
\end{itemize}

\subsection{Dataset Generation Details}
\label{sec: dataset-generation}
\begin{figure}[t]
  \centering
  \begin{subfigure}[t]{0.48\linewidth}
    \centering
    \includegraphics[width=\linewidth]{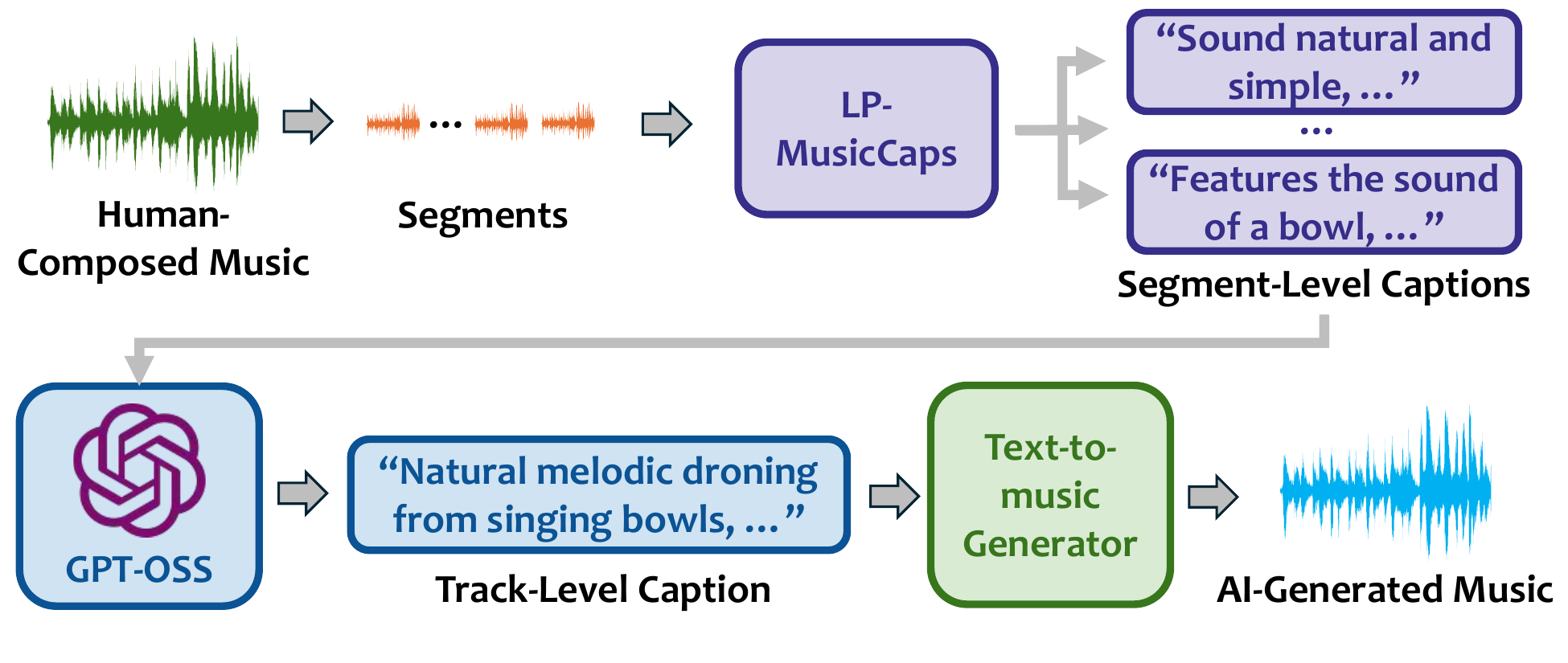}
    \caption{AI music generation.}
    \label{fig:captionpipeline}
  \end{subfigure}
  \hfill
  \begin{subfigure}[t]{0.51\linewidth}
    \centering
    \includegraphics[width=\linewidth]{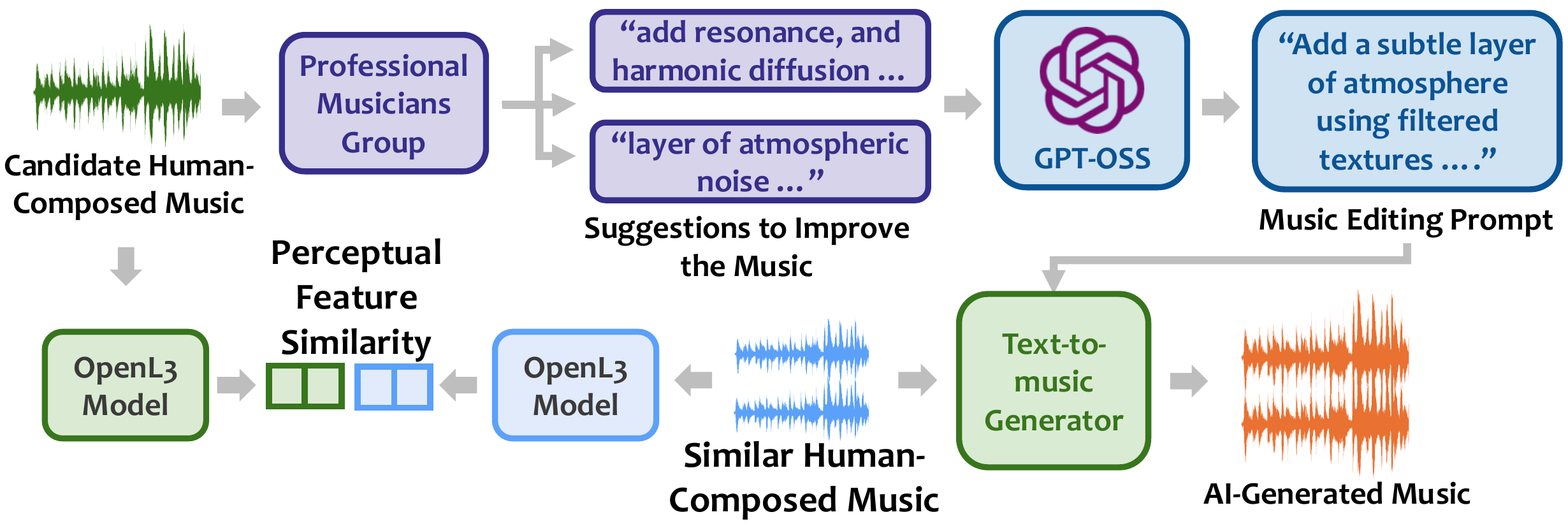}
    \caption{Human-AI co-creation.}
    \label{fig:co-creations}
  \end{subfigure}
  \vspace{-2mm}
  \caption{Pipelines for AI-generated and human-AI co-created music.}
  \label{fig:generation-pipelines}
  \vspace{-5mm}
\end{figure}

Figure~\ref{fig:captionpipeline} shows our music generation pipeline for stylistic alignment. Each human-composed track is divided into consecutive 10-second segments, which are individually captioned using LP-MusicCaps~\cite{doh2023lp}, a large-scale audio captioning model trained on over 500K music–tag pairs to describe musical attributes such as genre, instrumentation, rhythm, and mood. Because LP-MusicCaps is limited to clips shorter than 30 seconds, we employ GPT-OSS~\cite{agarwal2025gpt}, OpenAI's large language model with a 128K-token context window, to merge these segment-level captions into a coherent track-level description that preserves the overall structure and expressive intent. The resulting unified caption is then used as a text prompt for all ten text-to-music generators to produce stylistically matched AI compositions.
An example showing segment captions, the merged track-level description, and the system prompt used for GPT-OSS is provided in Appendix Table~\ref{tab:merged_caption_example}. This process ensures that both human and AI tracks share comparable musical content and stylistic attributes, enabling fair perceptual and algorithmic evaluations.

For open-source generators and Lyria RealTime, we use captions of human-composed tracks as prompts to generate 19,903 samples per model. For Suno and Udio, we reuse 10,000 tracks each from the recent SONICS dataset~\cite{rahman2025sonics}, which already includes high-quality vocal music generated by these commercial platforms based on the Genius Lyrics Dataset. Although SONICS does not employ our caption-based alignment pipeline, its songs originate from the same generators and composition sources, making them stylistically compatible with our benchmark. In total, we collected 179,224 AI-generated tracks.

\begin{table}[t]
  \centering
  \vspace{-1mm}
  \caption{Overview of \BenchName.}
  \label{tab:datasetoverview}
  \footnotesize

  % ---- table styling ----
  \rowcolors{2}{white}{lightblue} % alternating row colors
  \renewcommand{\arraystretch}{1.2} % increase row height
  
  \resizebox{0.85\linewidth}{!}{%
  \begin{tabular}{
      >{\centering\arraybackslash}m{7cm}
      >{\centering\arraybackslash}m{2cm}
      >{\centering\arraybackslash}m{2cm}
      >{\centering\arraybackslash}m{2cm}
  }
    \rowcolor{darkblue}
    \color{white}\textbf{Datasets / Models / Methods} &
    \color{white}\textbf{\# Tracks} &
    \color{white}\textbf{Total Hours} &
    \color{white}\textbf{Type} \\
    \multicolumn{4}{c}{\textbf{Human-Composed Music}} \\
    \midrule
    Genius-Lyrics~\cite{geniuslyrics} & 9,983 & 609 & Vocal \\
    JamendoMaxCaps~\cite{roy2025jamendomaxcaps} & 9,920 & 648 & Instrumental \\
    \midrule
    \multicolumn{4}{c}{\textbf{AI-Generated Music}} \\
    \midrule
    MusicGen~\cite{copet2023simple} & 19,903 & 165 & Instrumental \\
    AudioLDM2~\cite{liu2024audioldm2} & 19,903 & 165 & Instrumental \\
    Riffusion~\cite{Forsgren_Martiros_2022} & 19,903 & 169 & Vocal \\
    MAGNeT~\cite{ziv2024masked} & 19,903 & 165 & Instrumental \\
    JASCO~\cite{tal2024joint} & 19,903 & 55 & Instrumental \\
    Stable Audio Open~\cite{evans2025stable} & 19,903 & 165 & Instrumental \\
    Mustango~\cite{melechovsky2024mustango} & 19,903 & 56 & Vocal \\
    Lyria Realtime~\cite{lyriateam2025livemusicmodels} & 19,903 & 165 & Instrumental \\
    Suno~\cite{sunoai2023} & 10,000 & 485 & Vocal \\
    Udio~\cite{udio2024} & 10,000 & 307 & Vocal \\
    \midrule
    \multicolumn{4}{c}{\textbf{Human-AI Co-Created Music}} \\
    \midrule
    Co-Created Music & 9,887 & 82 & Vocal/Instr. \\
    \midrule
    \multicolumn{4}{c}{\textbf{Transformed Variants}} \\
    \midrule
    Lossy Compression (MP3 / OGG / AAC) & 53,805 & 571 & Vocal/Instr. \\
    Acoustic Post-Processing (EQ / Pitch-Time / Loudness) & 53,805 & 572 & Vocal/Instr. \\
    Additive Perturbations (Gaussian Noise / PGD~\cite{madry2017towards}) & 35,870 & 239 & Vocal/Instr. \\
    \midrule
    \multicolumn{4}{c}{\textbf{Out-Of-Domain}} \\
    \midrule
    Free Music Archive~\cite{defferrard2016fma} & 8,672 & 678 & Vocal/Instr. \\
    \multicolumn{4}{c}{\textbf{Total}} \\
    \midrule

        \textbf{\BenchName} & \textbf{361,166} & \textbf{5,296} & \textbf{Vocal/Instr.} \\
    \bottomrule
  \end{tabular}%
  }
\end{table}

\subsection{Curation Details of \BenchName}\label{sec:additional_dataset}
Unless otherwise specified, AI-generated tracks in \BenchName are produced with a maximum duration of 30s for all open-source models.
For models incapable of producing 30s outputs, we use their maximum supported length (e.g., 10s).
Transformations are applied only to the AI-generated music in the test set to ensure fair and consistent evaluation.
Below we describe the three groups of transformed variants.

\noindent\textbf{Lossy Compression.}
MP3 and AAC are encoded at 128~kbps, and OGG~Vorbis at quality level~3, introducing realistic yet perceptually mild distortions.

\noindent\textbf{Acoustic Post-processing.}
For \textit{equalization}, we applied six EQ strategies across three frequency ranges: Low-Boost/Low-Cut (25--250~Hz), corresponding to low-frequency instruments such as drums and bass; Mid-Boost/Mid-Cut (250~Hz--4~kHz), covering vocals, guitars, and keyboards; and High-Boost/High-Cut (4--20~kHz), affecting brightness and detail of percussive sounds. 
For each input track, one strategy is randomly selected. Cut and boost levels are set to $-6$~dB and $+6$~dB to preserve overall frequency balance. For tracks whose upper frequency bound does not reach 20~kHz, the three bands are proportionally calculated using the following ratios: $0–0.10$ (low), $0.10–0.35$ (mid), and $0.35–0.45$ (high). 
For \textit{pitch/time shifts}, each track is randomly transposed by ${-2,-1,+1,+2}$ semitones, and its playback speed set to either $0.9\times$ or $1.1\times$, altering pitch and timing while preserving naturalness.
For \textit{loudness normalization}, each track is adjusted to an integrated loudness of $-14$ LUFS, aligned with standards used by major streaming platforms, to ensure consistent playback levels while retaining dynamic range.

\noindent\textbf{Additive Perturbations.}
For \textit{Gaussian noise}, we inject zero-mean Gaussian noise with a standard deviation of 0.025, following prior work~\cite{ha2024organic}; this preserves perceptual quality while introducing measurable degradation to detector performance. For \textit{adversarial perturbations}, we follow the configuration described in the Appendix~\ref{sec:detectors} to first preprocess samples (to satisfy model input requirements) and craft PGD perturbations using gradients from two waveform-input models, M5 and RawNet2 because both operate on raw waveforms, the most fundamental audio representation, the resulting perturbations are more transferable to detectors with different input representations, yielding a more universal attack. During each PGD iteration, one of the two models is randomly selected to compute gradients, which are accumulated into the perturbation update. The attack is performed under the $\ell_\infty$ norm with step size $\alpha=0.001$, perturbation bound $\epsilon=0.005$, and $T=10$ iterations, balancing imperceptibility and attack effectiveness.

{
\tcbset{
  colback=white,                 
  colframe=darkblue,             
  colbacktitle=darkblue,         
  coltitle=white,                
  fonttitle=\bfseries,           
  arc=2mm,                       
  boxrule=0.6pt,                 
  left=2mm,right=2mm,top=1mm,bottom=1mm,
  enhanced,
}

\begin{table}[t]
\centering
\Large
\caption{System prompts for GPT-OSS and captions produced by the caption-based generation pipeline.}
\resizebox{0.75\linewidth}{!}{%
\begin{tabularx}{\textwidth}{X}
% --- System Prompt Box ---
\begin{tcolorbox}[title=GPT-OSS System Prompt]
You are a world-class expert in music captioning, specializing in merging multiple time-ordered segment captions into a single coherent and natural whole-song caption. Your task is to carefully integrate all important details from the segments - including instruments, vocals, mood, atmosphere, rhythm, style, and transitions - so that the final caption captures the essence of the entire song.
The merged caption must be concise yet comprehensive, written in fluent and descriptive language, suitable for music information retrieval or dataset annotation. It must cover the song as a whole rather than just isolated parts, and should avoid redundancy, contradictions, or missing key features. Output strictly $<=$ 300 characters in the 'final' channel, with no extra commentary or formatting.
\end{tcolorbox} \\[1mm]

% --- Segment Box ---
\begin{tcolorbox}[title=0s to 30s Segment-level Captions]
0s–10s: The song is an instrumental. The tempo is medium with a groovy bass line, keyboard accompaniment, steady drumming rhythm, and tambourine percussion. The song is a modern techno dance tune. The audio quality is poor.\\[1mm]
10s–20s: The song is an instrumental. The tempo is medium with a synthesizer arrangement, digital drumming rhythm, clapping percussion, and a strong bass line. The song is a techno dance rhythm. The audio quality is poor with ambient crowd noises.\\[1mm]
20s–30s: The low-quality recording features an electro song that consists of a punchy four-on-the-floor kick pattern, shimmering hi-hats, claps, and groovy synth bass. It sounds energetic and exciting.
\end{tcolorbox} \\[1mm]

% --- Merged Caption Box ---
\begin{tcolorbox}[title=Track-level Caption]
Groovy techno track: punchy four-on-the-floor kick, shimmering hi-hats, claps, and a lively synth lead with repetitive basslines. Mellow synth pads add depth, while a simple, syncopated rhythm drives the energy-ideal for club play.
\end{tcolorbox}
\end{tabularx}}
\label{tab:merged_caption_example}
\vspace{-6mm}
\end{table}
}

{
\tcbset{
  colback=white,                 
  colframe=darkblue,             
  colbacktitle=darkblue,         
  coltitle=white,                
  fonttitle=\bfseries,           
  arc=2mm,                       
  boxrule=0.6pt,                 
  left=2mm,right=2mm,top=1mm,bottom=1mm,
  enhanced,
}

\begin{table}[t]
\centering
\Large
\caption{Example of converting musicians' feedback into structured prompts.}
\resizebox{0.75\linewidth}{!}{%
\begin{tabularx}{\textwidth}{X}

% --- GPT-OSS Setting Box ---
\begin{tcolorbox}[title=GPT-OSS Role Prompt]
You are a music style prompt generator. Your task is to take short feedback about a music piece and rewrite it as a vivid, descriptive AI music generation prompt. The output should be detailed enough to guide a music generation model, describing instruments, style, mood, and overall tone.
\end{tcolorbox} \\[1mm]

% --- Musicians Feedback Box ---
\begin{tcolorbox}[title=Musicians Feedback]
Introduce a dreamlike ambience through spatial widening, gentle modulation, and dynamic blending of tonal layers to evoke surreal calmness.
\end{tcolorbox} \\[1mm]

% --- Structured Prompts Box ---
\begin{tcolorbox}[title=Structured Prompts]
A serene, dreamlike ambient composition characterized by wide spatial depth, gentle modulation, and fluid layering of tones. Soft, evolving textures merge seamlessly to create an atmosphere of surreal calm and emotional stillness, with smooth dynamics and subtle harmonic movement that evoke a sense of floating tranquility.
\end{tcolorbox}
\end{tabularx}}
\label{tab:co-created_example}
\end{table}

\vspace{-2mm}
\subsection{Example of Human-composed Caption}\label{sec:example_caption}
Table~\ref{tab:merged_caption_example} shows the system prompt used for GPT-OSS along with examples of segment-level and track-level captions. The first box defines GPT-OSS's role and instructs it to infer the overall musical style from multiple segments. The second box lists sample segment-level captions, each describing only a limited aspect of the piece. The final box presents the resulting track-level caption produced by GPT-OSS, which merges these fragments into a coherent description that accurately reflects the complete stylistic character of the human-composed music.

\vspace{-2mm}
\subsection{Example of Human-AI Co-Creation Prompts}\label{sec:example_cocreated}
% Table~\ref{tab:co-created_example} shows an example where GPT-OSS acts as a \textit{music style prompt constructor}, transforming musicians’ informal feedback into a structured, generation-ready prompt that preserves their intended stylistic refinements.
As illustrated in Appendix Figure~\ref{fig:co-creations}, the process begins with a set of candidate human-composed tracks evaluated by a panel of $31$ professional musicians, each providing qualitative suggestions (e.g., enhancing harmonic richness, adjusting rhythmic dynamics, or adding atmospheric elements). Because collecting expert feedback at scale is infeasible, we use the OpenL3 model~\cite{cramer2019look} to extract perceptual features from the reviewed tracks and identify additional human compositions with similar timbral and structural characteristics. These matched tracks are then paired with the corresponding expert feedback. The original natural-language comments are subsequently refined and standardized using GPT-OSS~\cite{agarwal2025gpt}, producing consistent music-editing prompts.

As shown in Table~\ref{tab:co-created_example}, we position GPT-OSS as a \textit{music style prompt constructor} by defining a focused role and clear output criteria, which ensures that its responses remain stylistically consistent and appropriately detailed. The second box presents an excerpt of musician feedback that points out specific aspects of the piece that could be strengthened. The final box shows how GPT-OSS restructures this brief and loosely organized feedback into a coherent prompt that retains the musicians’ intended refinements and presents them in a form that can directly guide music generation.

\vspace{-2mm}
\clearpage
\section{Detectors}
\label{sec: appendix-detectors}
\noindent\textbf{Open Research Baselines.} 
We benchmark a portfolio of detectors covering handcrafted features, convolutional encoders, and transformer architectures. These baselines span complementary input modalities (waveform vs. spectrogram) and inductive biases (local convolutional patterns vs. global attention), while following standardized, reproducible training configurations (Appendix~\ref{sec:detectors}).

\begin{itemize}[leftmargin=*, topsep=0pt, itemsep=0pt, parsep=0pt]

\item \textbf{Q\text{-}SVM}~\cite{singh2021qsvm} is a classical margin-based classifier trained on hand-crafted audio features, using a quadratic kernel to model simple nonlinear boundaries; a lightweight, interpretable baseline.
\item \textbf{Deezer}~\cite{afchar2025fourier} instroduces a theory-driven AI music detector grounded in Fourier analysis of deconvolution layers: it mathematically proves that strided deconvolution operations in generative architectures inherently periodize the spectrum, producing distinctive spectral peaks independent of training data; a simple logistic regressor on the resulting frequency fingerprint then achieves detection accuracy on par with deep learning baselines; the \emph{best paper} at ISMIR 2025.
\item \textbf{M5}~\cite{dai2017very} is a lightweight CNN operating directly on raw waveforms, which serves as a proof-of-concept model to assess whether the detection task is tractable using minimal inductive bias.
\item \textbf{RawNet2}~\cite{tak2021rawnet2} is an end-to-end waveform model originally developed for speaker verification and anti-spoofing, which combines sinc-based convolutional layers, residual blocks, and gated recurrent units to extract temporal features from raw waveforms.
\item \textbf{ConvNeXt}~\cite{liu2022convnet} is a convolutional architecture with strong local inductive bias, using stagewise downsampling and residual blocks for high-capacity modeling of mel-spectrogram inputs.
\item \textbf{ResNet18+Spec}~\cite{he2016deep} is a residual CNN trained on log-Mel spectrograms, using BasicBlocks and skip connections to capture mid- to long-range time–frequency structure; with strided downsampling, batch normalization, and average pooling + linear head.
\item \textbf{ViT}~\cite{dosovitskiy2021vit} is a patch-based Transformer on Mel spectrograms that projects spectrogram patches into tokens and models global context with self-attention, which offers strong long range structure modeling while incurring token growth with input duration.
\item \textbf{AST}~\cite{gong2021ast} adapts ViT to audio by patchifying log-Mel spectrograms into time–frequency segments. Its self-attention captures long-range dependencies and cross-band interactions, and pretrained backbones transfer to music information retrieval tasks.
\item \textbf{SpecTTTra}~\cite{rahman2025sonics} is a spectrogram Transformer detector that reduces token count via spectrotemporal tokenization and applies global self-attention over compact tokens.
\end{itemize}
\noindent\textbf{Commercial Detectors.} We leverage two widely used commercial detectors, \textbf{Ircam Amplify}~\cite{ircamamplifyMusicDetector} and \textbf{Resemble AI}~\cite{resembleDetectDeepfakes}. Both systems are treated as black boxes: we query their official APIs and log the returned detection scores and labels for subsequent analysis. While Resemble AI is primarily a voice deepfake detector, its capabilities extend to AI-generated music detection.
\begin{table}
  \centering
  \caption{Machine-based detectors evaluated. Type: open-source (OS) or commercial (COM).}
  \label{tab:detectors}
  \scriptsize  % smaller than \footnotesize

  \rowcolors{2}{white}{lightblue} % alternating row colors
  \renewcommand{\arraystretch}{1.05} % slightly reduce row height
  
  % make the whole table a bit narrower and let columns shrink
  \resizebox{0.85\linewidth}{!}{%
  \begin{tabular}{>{\centering\arraybackslash}m{2.1cm} 
                  >{\centering\arraybackslash}m{0.8cm} 
                  >{\centering\arraybackslash}m{1.7cm} 
                  >{\centering\arraybackslash}m{3.0cm} 
                  >{\centering\arraybackslash}m{1.7cm}}
    \rowcolor{darkblue}
    \color{white}\textbf{Detector} & 
    \color{white}\textbf{Type} & 
    \color{white}\textbf{Backbone} & 
    \color{white}\textbf{Input} & 
    \color{white}\textbf{\#Params} \\
    Q-SVM~\cite{singh2021qsvm}         & OS  & Classical   & Handcrafted Features  & 0.2M            \\
        Deezer~\cite{afchar2025fourier}         & OS  & Classical   & Handcrafted Features  & 0.004M            \\
    M5~\cite{dai2017very}             & OS  & CNN         & Waveform     & 0.4M         \\
    RawNet2~\cite{tak2021rawnet2}        & OS  & CNN         & Waveform     & 8M       \\
    ConvNeXt~\cite{liu2022convnet}       & OS  & CNN         & Mel Spec     & 28M      \\
    ResNet18+Spec~\cite{he2016deep}  & OS  & CNN         & Log Spec  & 11M       \\
    AST~\cite{gong2021ast}            & OS  & Transformer & Mel Fbank     & 88M      \\
    ViT~\cite{dosovitskiy2021vit}            & OS  & Transformer & Mel Spec     & 17M      \\
    SpecTTTra~\cite{rahman2025sonics}      & OS  & Transformer & Mel Spec  & 17M       \\
    Ircam Amplify~\cite{ircamamplifyMusicDetector}  & COM & Black-box API & ---         & ---            \\
    Resemble AI~\cite{resembleDetectDeepfakes}    & COM & Black-box API & ---         & ---            \\
    \bottomrule
  \end{tabular}%
  }
  \vspace{-4mm}
\end{table}

\subsection{Details of Detector Architectures and Training Configurations}\label{sec:detectors}
Each music track contributes a single, randomly selected 10s mono excerpt sampled at 48kHz. Segments are cropped deterministically from the annotated start time, and normalized whenever non-silent. Spectrogram-based detectors operate on either log-magnitude STFTs or log-mel representations after mean–variance normalization. Moreover, to mitigate class imbalance, we oversample the minority class via simple replication; for each replicated track, we also draw a new randomly selected 10s excerpt to reduce overfitting while preserving label balance. Alternative approaches, including class-weighted losses, were considered but consistently yielded inferior performance.
All detectors are trained with batches of 32 and early stopping based on validation loss with patience of 10.

During training, we optimize M5, RawNet2, and ResNet18+Spec models with Adam (initial learning rate $1\times10^{-4}$, weight decay $1\times10^{-4}$), clip gradient norms at 5, and decay the learning rate by a factor of 0.1 every 20 epochs.
The AST detector uses the ViT base-384 backbone pretrained on ImageNet. We employ binary cross-entropy with logits, Adam with $\beta=(0.95,0.999)$ and weight decay $5\times10^{-7}$, and warm up the learning rate linearly during the first 1,000 steps before applying a multi-step decay beginning at epoch 10. Mixup is applied with probability $0.5$, together with SpecAugment (time and frequency masks), on the $128\times 1000$ log-mel spectrograms.

Additionally, ConvNeXt, SpecTTTra, and ViT share a unified pipeline: waveforms are converted to normalized log-mel spectrograms, augmented with mixup ($\alpha{=}2.5$, $p{=}0.5$) and SpecAugment (two time and one frequency mask), then bilinearly resized to $128{\times}128$ before encoding. Training uses AdamW (lr=$8{\times}10^{-4}$, weight decay=$5{\times}10^{-2}$) with a cosine schedule, five warm-up epochs, and gradient clipping at~5. SpecTTTra adopts the $\alpha$ variant ($t{=}3,f{=}1$) for improved accuracy.
%Additionally, ConvNeXt, SpecTTTra and ViT share a unified pipeline. Waveforms are mapped to normalized log-mel spectrograms, then augmented with mixup ($\alpha=2.5$, probability $0.5$) and SpecAugment (two time masks and one frequency mask). The resulting features are bilinearly resized to $128\times128$ before being fed into the encoder. Training uses AdamW with a base learning rate of $8\times10^{-4}$, weight decay of $5\times10^{-2}$), a cosine learning rate schedule with five warm-up epochs, and gradient clipping at 5. For SpecTTTra, we use the $\alpha$ variant ($t{=}3,f{=}1$) to prioritize accuracy.

For classical baseline, Q-SVM, we represent each 10s clip with a 14-dimensional descriptor comprising the first four central moments of the bispectrum magnitude, the corresponding four moments of the bispectrum phase, and the mean/variance of mel-frequency cepstral coefficients together with their first- and second-order temporal differences. Feature vectors are standardized using training-set statistics and evaluated with support-vector machines employing a quadratic (degree-2 polynomial) kernel.
Hyperparameters $(C,\gamma)$ are selected via grid search inside a stratified five-fold cross-validation routine.

\vspace{-2mm}
\section{Experiment Results} 
\subsection{Overall Machine Capabilities} 
\label{sec: results}
\begin{table*}[t]
\centering
\caption{Default-setting results: Overall and per-generator performance (\%).}
\vspace{-2mm}
\label{tab:in_domain}
\scriptsize
\setlength{\tabcolsep}{2.8pt}
\renewcommand{\arraystretch}{1.08}
\resizebox{\linewidth}{!}{%
\begin{tabular}{l|ccccc|cccccccccc}
\toprule
& \multicolumn{5}{c|}{\textbf{Overall}}
& \multicolumn{10}{c}{\textbf{Per-Generator FNRs}} \\
\cmidrule(lr){2-6}
\cmidrule(lr){7-16}
\textbf{Detector}
& $\mathbf{ACC}_{\mathbf{B}}$
& \textbf{Prec.}
& \textbf{Rec.}
& \textbf{FPR}
& \textbf{FNR}
& \textbf{MusicGen}
& \textbf{AudioLDM2}
& \textbf{Riffusion}
& \textbf{MAGNeT}
& \textbf{JASCO}
& \makecell{\textbf{Stable}\\\textbf{Audio}}
& \textbf{Mustango}
& \makecell{\textbf{Lyria}\\\textbf{RT}}
& \textbf{Suno}
& \textbf{Udio} \\
\midrule
\rowcolor{gray!15}
\multicolumn{16}{l}{\textbf{Open-source}} \\
Q-SVM & \cellcolor{softblue!20}68.9 & \cellcolor{softblue!70}95.6 & \cellcolor{softblue!18}65.3 & \cellcolor{softblue!50}27.4 & \cellcolor{softblue!42}34.7 & \cellcolor{softblue!28}53.6 & \cellcolor{softblue!78}7.4 & \cellcolor{softblue!64}20.2 & \cellcolor{softblue!42}35.3 & \cellcolor{softblue!34}45.2 & \cellcolor{softblue!26}57.8 & \cellcolor{softblue!74}10.6 & \cellcolor{softblue!36}42.5 & \cellcolor{softblue!40}39.1 & \cellcolor{softblue!38}40.7 \\
Deezer & \cellcolor{softblue!78}97.6 & \cellcolor{softblue!88}99.8 & \cellcolor{softblue!74}97.2 & \cellcolor{softblue!86}2.0 & \cellcolor{softblue!82}2.8 & \cellcolor{softblue!90}0.3 & \cellcolor{softblue!90}0.3 & \cellcolor{softblue!82}3.6 & \cellcolor{softblue!90}0.1 & \cellcolor{softblue!88}0.9 & \cellcolor{softblue!84}1.8 & \cellcolor{softblue!90}0.1 & \cellcolor{softblue!68}15.6 & \cellcolor{softblue!86}1.4 & \cellcolor{softblue!82}3.6 \\
M5 & \cellcolor{softblue!68}94.3 & \cellcolor{softblue!78}98.9 & \cellcolor{softblue!78}98.6 & \cellcolor{softblue!74}9.9 & \cellcolor{softblue!86}1.4 & \cellcolor{softblue!88}0.7 & \cellcolor{softblue!90}0.2 & \cellcolor{softblue!90}0.2 & \cellcolor{softblue!90}0.2 & \cellcolor{softblue!86}1.6 & \cellcolor{softblue!82}3.3 & \cellcolor{softblue!88}0.3 & \cellcolor{softblue!82}3.0 & \cellcolor{softblue!86}1.7 & \cellcolor{softblue!78}4.8 \\
RawNet2 & \cellcolor{softblue!42}82.3 & \cellcolor{softblue!72}96.5 & \cellcolor{softblue!72}96.2 & \cellcolor{softblue!44}31.6 & \cellcolor{softblue!80}3.8 & \cellcolor{softblue!76}6.4 & \cellcolor{softblue!88}0.6 & \cellcolor{softblue!86}1.3 & \cellcolor{softblue!86}1.3 & \cellcolor{softblue!78}5.3 & \cellcolor{softblue!70}12.9 & \cellcolor{softblue!86}1.3 & \cellcolor{softblue!82}3.0 & \cellcolor{softblue!84}2.1 & \cellcolor{softblue!82}3.3 \\
ConvNeXt & \cellcolor{softblue!88}99.8 & \cellcolor{softblue!90}\textbf{100.0} & \cellcolor{softblue!88}99.8 & \cellcolor{softblue!90}0.2 & \cellcolor{softblue!90}0.2 & \cellcolor{softblue!90}0.1 & \cellcolor{softblue!90}\textbf{0.0} & \cellcolor{softblue!90}0.1 & \cellcolor{softblue!90}0.1 & \cellcolor{softblue!88}0.7 & \cellcolor{softblue!88}0.8 & \cellcolor{softblue!90}0.1 & \cellcolor{softblue!90}0.2 & \cellcolor{softblue!88}0.4 & \cellcolor{softblue!90}0.1 \\
ResNet18+Spec & \cellcolor{softblue!90}\textbf{100.0} & \cellcolor{softblue!90}\textbf{100.0} & \cellcolor{softblue!90}\textbf{100.0} & \cellcolor{softblue!90}\textbf{0.0} & \cellcolor{softblue!90}\textbf{0.0} & \cellcolor{softblue!90}\textbf{0.0} & \cellcolor{softblue!90}\textbf{0.0} & \cellcolor{softblue!90}\textbf{0.0} & \cellcolor{softblue!90}\textbf{0.0} & \cellcolor{softblue!90}\textbf{0.0} & \cellcolor{softblue!90}0.1 & \cellcolor{softblue!90}\textbf{0.0} & \cellcolor{softblue!90}\textbf{0.0} & \cellcolor{softblue!90}0.1 & \cellcolor{softblue!90}\textbf{0.0} \\
AST & \cellcolor{softblue!90}\textbf{100.0} & \cellcolor{softblue!90}\textbf{100.0} & \cellcolor{softblue!90}\textbf{100.0} & \cellcolor{softblue!90}\textbf{0.0} & \cellcolor{softblue!90}\textbf{0.0} & \cellcolor{softblue!90}\textbf{0.0} & \cellcolor{softblue!90}\textbf{0.0} & \cellcolor{softblue!90}\textbf{0.0} & \cellcolor{softblue!90}\textbf{0.0} & \cellcolor{softblue!90}\textbf{0.0} & \cellcolor{softblue!90}\textbf{0.0} & \cellcolor{softblue!90}\textbf{0.0} & \cellcolor{softblue!90}\textbf{0.0} & \cellcolor{softblue!90}0.1 & \cellcolor{softblue!90}\textbf{0.0} \\
ViT & \cellcolor{softblue!86}99.7 & \cellcolor{softblue!90}\textbf{100.0} & \cellcolor{softblue!86}99.7 & \cellcolor{softblue!90}0.3 & \cellcolor{softblue!90}0.3 & \cellcolor{softblue!90}\textbf{0.0} & \cellcolor{softblue!90}\textbf{0.0} & \cellcolor{softblue!90}\textbf{0.0} & \cellcolor{softblue!90}\textbf{0.0} & \cellcolor{softblue!90}0.3 & \cellcolor{softblue!84}1.6 & \cellcolor{softblue!90}\textbf{0.0} & \cellcolor{softblue!90}0.3 & \cellcolor{softblue!90}0.2 & \cellcolor{softblue!90}0.1 \\
SpecTTTra & \cellcolor{softblue!82}99.5 & \cellcolor{softblue!88}99.9 & \cellcolor{softblue!82}99.5 & \cellcolor{softblue!88}0.6 & \cellcolor{softblue!88}0.5 & \cellcolor{softblue!88}0.5 & \cellcolor{softblue!90}0.1 & \cellcolor{softblue!90}0.1 & \cellcolor{softblue!88}0.5 & \cellcolor{softblue!86}1.2 & \cellcolor{softblue!84}1.5 & \cellcolor{softblue!90}0.2 & \cellcolor{softblue!90}0.3 & \cellcolor{softblue!88}0.6 & \cellcolor{softblue!90}0.3 \\
\rowcolor{gray!15}
\multicolumn{16}{l}{\textbf{Commercial}} \\
Ircam Amplify & \cellcolor{softblue!40}81.0 & \cellcolor{softblue!78}98.6 & \cellcolor{softblue!24}71.1 & \cellcolor{softblue!76}9.0 & \cellcolor{softblue!48}28.9 & \cellcolor{softblue!90}\textbf{0.0} & \cellcolor{softblue!78}7.4 & \cellcolor{softblue!5}93.2 & \cellcolor{softblue!80}4.2 & \cellcolor{softblue!78}7.4 & \cellcolor{softblue!32}46.2 & \cellcolor{softblue!76}8.2 & \cellcolor{softblue!10}79.2 & \cellcolor{softblue!90}\textbf{0.0} & \cellcolor{softblue!46}30.4 \\
Resemble.AI & \cellcolor{softblue!5}46.9 & \cellcolor{softblue!52}89.2 & \cellcolor{softblue!18}64.7 & \cellcolor{softblue!16}70.8 & \cellcolor{softblue!42}35.3 & \cellcolor{softblue!62}21.2 & \cellcolor{softblue!76}9.4 & \cellcolor{softblue!40}38.4 & \cellcolor{softblue!56}24.8 & \cellcolor{softblue!56}25.2 & \cellcolor{softblue!38}40.8 & \cellcolor{softblue!44}33.4 & \cellcolor{softblue!34}43.6 & \cellcolor{softblue!8}85.2 & \cellcolor{softblue!12}77.2 \\
\bottomrule
\end{tabular}%
}
\vspace{-4mm}
\end{table*}

We begin by examining the general performance of all machine-based detectors under the default setting. From Table~\ref{tab:in_domain}, We can observe that convolutional and transformer-based spectrogram models achieve near-perfect performance. ResNet18+Spec and AST are nearly perfect on $\mathrm{ACC_B}$, precision, and recall with zero FPR/FNR. ConvNeXt, ViT, and SpecTTTra follow closely, with overall balanced accuracies between $99.5\%$ and $99.8\%$, differing only by rounding error, and correspondingly negligible FPR and FNR values.
Waveform CNNs also perform strongly but fall slightly behind this group. M5 achieves balanced accuracy of $94.3\%$ (precision $98.9\%$, recall $98.6\%$, $\mathrm{FNR}=1.4\%$), whereas RawNet2 is limited by a higher $\mathrm{FPR}$ of $31.6\%$ ($\mathrm{ACC_B}=82.3\%$). Deezer’s handcrafted feature-based detector performs robustly ($\mathrm{ACC_B}=97.6\%$), while Q-SVM shows substantially weaker discrimination ability ($\mathrm{ACC_B}=68.9\%$).
Among commercial detectors, performance varies more widely. Ircam Amplify maintains high precision ($98.6\%$) but lower recall ($71.1\%$; $\mathrm{FNR}=28.9\%$), while Resemble AI exhibits the weakest performance overall ($\mathrm{ACC_B}=46.9\%$, $\mathrm{FPR}=70.8\%$), frequently misclassifying human audio as AI. These results suggest that Ircam Amplify adopts a conservative, precision-oriented design, whereas Resemble AI appears more prone to false alarms and domain sensitivity.
Commercial APIs underperform, likely due to a distribution gap between their proprietary training data and the test set in \BenchName, which spans 10 generators and includes both instrumental and vocal music. 

\vspace{-3mm}
\subsubsection{Performance Consistency Across Generators}
\begin{sloppypar}
Across the default evaluation from Table~\ref{tab:in_domain}, spectrogram-based models form a clear top tier. ResNet18+Spec, AST, ConvNeXt, ViT, and SpecTTTra are effectively perfect on all ten generators (FNRs at or below $1.6\%$), indicating that time–frequency representations provide sufficient cues for separation regardless of backbone. Among waveform models, M5 is uniformly strong, staying below $5\%$ on every generator and peaking at $0.2\%$ on AudioLDM2/Riffusion/MAGNeT, while RawNet2 is slightly weaker but steady overall ($0.6\%\text{--}12.9\%$, highest on Stable Audio Open). Deezer's detector performs strongly overall, but is less stable on Lyria Realtime, where its FNR rises to $15.6\%$. In contrast, Q-SVM remains unstable: it nears strong baselines on AudioLDM2 ($7.4\%$) and Mustango ($10.6\%$) but drops on Riffusion ($20.2\%$) and more sharply on Suno/Udio ($39.1\%$/$40.7\%$) and Stable Audio Open ($57.8\%$). Commercial detectors are weaker and less consistent. Ircam Amplify is near-perfect on MusicGen/Suno ($0.0\%$/$0.0\%$) and strong on MAGNeT ($4.2\%$) yet collapses on Riffusion/Lyria Realtime ($93.2\%$/$79.2\%$) and is middling on newer generators like Stable Audio Open/Udio ($46.2\%$/$30.4\%$). Resemble AI shows the reverse pattern, stronger on AudioLDM2 ($9.4\%$) but poor on Suno/Udio ($85.2\%$/$77.2\%$).
The results suggest that spectrogram-based architectures generalize reliably across diverse generation methods, whereas the classical AI detector or commercial APIs remain sensitive to generator-specific artifacts.
\end{sloppypar}
% \subsection{Impact of Training Set Selection}
\subsection{Detailed Results of Leave-One-Model-Out}
\label{sec: leave-results}
\begin{table*}[!t]
\centering
\caption{Leave-One-Model-Out setting results (\%). Each cell reports ACC$_B$ on the first line and FPR / FNR on the second line.}
\vspace{-2mm}
\label{tab:leave_one_out}
\begingroup
\scriptsize
\setlength{\tabcolsep}{3.4pt}
\resizebox{\linewidth}{!}{%
\renewcommand{\arraystretch}{1.10}
\begin{NiceTabular}{c|ccccccccc}[colortbl-like]
\CodeBefore
    \rowcolor{gray!15}{1,2}
    \arrayrulecolor{black}
\Body
\toprule
\Block{2-1}{\textbf{Left-out}\\\textbf{Model}}
& \Block{1-1}{\textbf{Q-SVM}}
& \Block{1-1}{\textbf{Deezer}}
& \Block{1-1}{\textbf{M5}}
& \Block{1-1}{\textbf{RawNet2}}
& \Block{1-1}{\textbf{ConvNeXt}}
& \Block{1-1}{\makecell{\textbf{ResNet18+Spec}}}
& \Block{1-1}{\textbf{AST}}
& \Block{1-1}{\textbf{ViT}}
& \Block{1-1}{\textbf{SpecTTTra}} \\
&
\makecell{\scriptsize ACC$_B$\\\scriptsize FPR / FNR}
& \makecell{\scriptsize ACC$_B$\\\scriptsize FPR / FNR}
& \makecell{\scriptsize ACC$_B$\\\scriptsize FPR / FNR}
& \makecell{\scriptsize ACC$_B$\\\scriptsize FPR / FNR}
& \makecell{\scriptsize ACC$_B$\\\scriptsize FPR / FNR}
& \makecell{\scriptsize ACC$_B$\\\scriptsize FPR / FNR}
& \makecell{\scriptsize ACC$_B$\\\scriptsize FPR / FNR}
& \makecell{\scriptsize ACC$_B$\\\scriptsize FPR / FNR}
& \makecell{\scriptsize ACC$_B$\\\scriptsize FPR / FNR} \\
\midrule

MusicGen
& \LOMOCell{27}{59.0}{34.7 / 47.2}
& \LOMOCell{85}{97.2}{2.1 / 3.4}
& \LOMOCell{79}{93.2}{2.5 / 11.0}
& \LOMOCell{61}{81.5}{12.4 / 24.5}
& \LOMOCell{84}{95.8}{1.0 / 7.5}
& \LOMOCellB{90}{100.0}{0.0 / 0.0}
& \LOMOCellB{90}{100.0}{0.1 / 0.0}
& \LOMOCell{88}{98.4}{0.5 / 2.8}
& \LOMOCell{86}{97.5}{0.7 / 4.3} \\

AudioLDM2
& \LOMOCell{57}{78.6}{37.6 / 5.2}
& \LOMOCell{75}{91.0}{2.2 / 15.9}
& \LOMOCell{46}{71.3}{3.5 / 53.9}
& \LOMOCell{71}{87.4}{16.5 / 8.8}
& \LOMOCell{82}{94.7}{1.0 / 9.6}
& \LOMOCellB{90}{100.0}{0.0 / 0.0}
& \LOMOCellB{90}{100.0}{0.0 / 0.0}
& \LOMOCell{67}{85.0}{0.4 / 29.6}
& \LOMOCell{54}{76.6}{0.6 / 46.2} \\

Riffusion
& \LOMOCell{48}{72.5}{36.3 / 18.7}
& \LOMOCell{10}{51.1}{1.2 / 96.6}
& \LOMOCell{14}{50.8}{2.5 / 96.0}
& \LOMOCell{11}{48.8}{14.5 / 87.8}
& \LOMOCell{83}{95.6}{1.0 / 7.9}
& \LOMOCellB{90}{100.0}{0.0 / 0.0}
& \LOMOCell{90}{99.9}{0.1 / 0.0}
& \LOMOCell{50}{74.2}{0.3 / 51.3}
& \LOMOCell{50}{73.8}{0.5 / 51.8} \\

MAGNeT
& \LOMOCell{37}{65.7}{33.5 / 35.2}
& \LOMOCell{89}{98.7}{2.1 / 0.5}
& \LOMOCell{87}{98.1}{2.3 / 1.6}
& \LOMOCell{71}{87.5}{22.1 / 3.0}
& \LOMOCell{88}{98.6}{0.6 / 2.2}
& \LOMOCellB{90}{100.0}{0.0 / 0.0}
& \LOMOCellB{90}{100.0}{0.0 / 0.0}
& \LOMOCell{88}{98.7}{0.3 / 2.4}
& \LOMOCell{89}{99.2}{0.3 / 1.2} \\

JASCO
& \LOMOCell{27}{59.2}{39.1 / 42.5}
& \LOMOCell{82}{95.0}{2.2 / 7.9}
& \LOMOCell{64}{83.5}{16.6 / 16.4}
& \LOMOCell{65}{84.0}{19.5 / 12.5}
& \LOMOCell{81}{94.1}{0.9 / 10.9}
& \LOMOCellB{90}{100.0}{0.0 / 0.0}
& \LOMOCellB{90}{100.0}{0.0 / 0.0}
& \LOMOCell{86}{97.2}{0.3 / 5.3}
& \LOMOCell{82}{94.9}{0.3 / 10.0} \\

\makecell{Stable\\Audio Open}
& \LOMOCell{7}{46.6}{32.9 / 73.9}
& \LOMOCell{12}{52.5}{2.4 / 92.7}
& \LOMOCell{15}{51.7}{2.1 / 94.5}
& \LOMOCell{33}{63.1}{10.2 / 63.5}
& \LOMOCell{59}{80.0}{0.8 / 39.2}
& \LOMOCellB{90}{99.7}{0.0 / 0.5}
& \LOMOCell{89}{99.4}{0.0 / 1.2}
& \LOMOCell{73}{89.2}{0.1 / 21.5}
& \LOMOCell{75}{90.1}{0.1 / 19.7} \\

Mustango
& \LOMOCell{48}{72.8}{37.4 / 16.9}
& \LOMOCell{89}{98.8}{2.2 / 0.3}
& \LOMOCell{86}{97.2}{3.5 / 2.0}
& \LOMOCell{77}{91.7}{11.6 / 5.1}
& \LOMOCell{87}{98.2}{1.1 / 2.4}
& \LOMOCellB{90}{100.0}{0.0 / 0.0}
& \LOMOCellB{90}{100.0}{0.0 / 0.0}
& \LOMOCell{88}{98.8}{0.4 / 2.1}
& \LOMOCell{88}{98.5}{0.4 / 2.7} \\

\makecell{Lyria\\Realtime}
& \LOMOCell{26}{58.8}{32.6 / 49.7}
& \LOMOCell{10}{51.0}{0.6 / 97.3}
& \LOMOCell{57}{78.9}{1.4 / 40.9}
& \LOMOCell{42}{68.8}{15.0 / 47.4}
& \LOMOCell{58}{79.5}{1.0 / 40.1}
& \LOMOCellB{90}{100.0}{0.0 / 0.0}
& \LOMOCell{84}{96.3}{0.0 / 7.4}
& \LOMOCell{66}{84.6}{0.3 / 30.5}
& \LOMOCell{66}{84.6}{0.4 / 30.5} \\

Suno
& \LOMOCell{39}{66.9}{39.1 / 27.0}
& \LOMOCell{88}{98.1}{2.1 / 1.7}
& \LOMOCell{85}{97.0}{3.9 / 2.0}
& \LOMOCell{69}{86.3}{22.2 / 5.3}
& \LOMOCell{88}{98.6}{1.6 / 1.2}
& \LOMOCell{90}{99.8}{0.1 / 0.4}
& \LOMOCellB{90}{99.9}{0.0 / 0.3}
& \LOMOCell{89}{99.2}{0.4 / 1.2}
& \LOMOCell{89}{99.1}{0.9 / 0.8} \\

Udio
& \LOMOCell{25}{58.2}{41.3 / 42.4}
& \LOMOCell{38}{68.3}{2.2 / 61.3}
& \LOMOCell{79}{93.1}{2.7 / 11.1}
& \LOMOCell{65}{83.6}{21.1 / 11.7}
& \LOMOCell{87}{97.8}{3.4 / 1.1}
& \LOMOCell{89}{99.6}{0.0 / 0.9}
& \LOMOCellB{90}{99.9}{0.0 / 0.2}
& \LOMOCell{88}{99.0}{0.2 / 1.9}
& \LOMOCell{88}{98.7}{0.8 / 1.9} \\

\bottomrule
\end{NiceTabular}
}
\endgroup
\vspace{-2mm}
\end{table*}

\begin{table*}[!t]
\centering
\caption{OOD accuracy of automatic detectors.}
\vspace{-2mm}
\label{tab:ood_accuracy}
\begingroup
\setlength{\tabcolsep}{5pt}
\resizebox{0.82\linewidth}{!}{%
\renewcommand{\arraystretch}{1.0}
\begin{NiceTabular}{l|ccccccccc}[colortbl-like]
\CodeBefore
    \rowcolor{gray!15}{1}
    \arrayrulecolor{black}
\Body
\toprule
\textbf{Metric}
& \textbf{Q-SVM}
& \textbf{Deezer} 
& \textbf{M5}
& \textbf{RawNet2}
& \textbf{ConvNeXt}
& \makecell{\textbf{ResNet18}\\\textbf{+Spec}}
& \textbf{AST}
& \textbf{ViT}
& \textbf{SpecTTTra}
\\
\midrule

\textbf{Accuracy}
& \cellcolor{softblue!28}55.9
& \cellcolor{softblue!90}\textbf{94.4} 
& \cellcolor{softblue!58}75.8
& \cellcolor{softblue!10}39.6
& \cellcolor{softblue!14}43.5
& \cellcolor{softblue!5}28.4
& \cellcolor{softblue!64}79.8
& \cellcolor{softblue!12}40.7
& \cellcolor{softblue!10}39.7
\\

\bottomrule
\end{NiceTabular}
}
\endgroup
\vspace{-2mm}
\end{table*}

To more rigorously assess model generalizability, we adopt a Leave-One-Model-Out protocol: in each round, one generator is excluded entirely from training, and the detector is then evaluated on the music produced by that unseen generator, along with human compositions, to measure both false positive and false negative rates. This setting provides a more realistic test of cross-generator transfer and exposes how well each detector generalizes beyond the distributions it was trained on.
The detailed results are shown in Table~\ref{tab:leave_one_out}.

\subsection{Performance of Machine-based Detectors on 30-second Music Segments}
\label{subsec:30s}
\begin{table*}[t]
\centering
\caption{Performance of automatic detectors using 30-second clips.}
\label{tab:30s_results}
\begingroup
\setlength{\tabcolsep}{5pt}
\resizebox{0.95\linewidth}{!}{%
\renewcommand{\arraystretch}{1.0}
\begin{NiceTabular}{l|ccccc|cccccccc}[colortbl-like]
\CodeBefore
    \rowcolor{gray!15}{1,2}
    \arrayrulecolor{black}
\Body
\toprule
\Block{2-1}{\makecell[c]{\textbf{Automatic}\\\textbf{Detector}}} &
\Block{1-5}{\textbf{Overall}} & & & & &
\Block{1-8}{\textbf{Per-Generator FNRs}} \\
& 
$\mathbf{ACC}_{\mathbf{B}}$ &
\textbf{Precision} &
\textbf{Recall} &
\textbf{FPR} &
\textbf{FNR} &
\textbf{Suno} &
\textbf{MAGNeT} &
\textbf{MusicGen} &
\textbf{AudioLDM2} &
\textbf{Riffusion} &
\textbf{Udio} &
\makecell{\textbf{Lyria}\\\textbf{Realtime}} &
\makecell{\textbf{Stable}\\\textbf{Audio Open}} \\
\midrule

Q-SVM
& \cellcolor{softblue!20}76.9
& \cellcolor{softblue!52}96.5
& \cellcolor{softblue!16}72.0
& \cellcolor{softblue!7}18.2
& \cellcolor{softblue!5}28.0
& \cellcolor{softblue!5}71.5
& \cellcolor{softblue!5}41.8
& \cellcolor{softblue!5}44.5
& \cellcolor{softblue!5}31.1
& \cellcolor{softblue!18}5.7
& \cellcolor{softblue!10}13.5
& \cellcolor{softblue!12}11.7
& \cellcolor{softblue!7}18.5 \\

Deezer
& \cellcolor{softblue!74}99.06
& \cellcolor{softblue!88}99.91
& \cellcolor{softblue!68}98.78
& \cellcolor{softblue!64}0.66
& \cellcolor{softblue!52}1.22
& \cellcolor{softblue!66}0.60
& \cellcolor{softblue!90}\textbf{0.00}
& \cellcolor{softblue!88}0.05
& \cellcolor{softblue!84}0.15
& \cellcolor{softblue!54}1.00
& \cellcolor{softblue!46}1.30
& \cellcolor{softblue!20}5.02
& \cellcolor{softblue!44}1.40 \\

M5
& \cellcolor{softblue!64}98.3
& \cellcolor{softblue!78}99.7
& \cellcolor{softblue!70}98.9
& \cellcolor{softblue!34}2.3
& \cellcolor{softblue!50}1.1
& \cellcolor{softblue!62}0.7
& \cellcolor{softblue!78}0.3
& \cellcolor{softblue!66}0.6
& \cellcolor{softblue!78}0.3
& \cellcolor{softblue!82}0.2
& \cellcolor{softblue!44}1.4
& \cellcolor{softblue!32}2.5
& \cellcolor{softblue!30}2.7 \\

RawNet2
& \cellcolor{softblue!42}93.1
& \cellcolor{softblue!66}98.5
& \cellcolor{softblue!56}97.0
& \cellcolor{softblue!12}10.7
& \cellcolor{softblue!28}3.0
& \cellcolor{softblue!58}0.8
& \cellcolor{softblue!37}2.0
& \cellcolor{softblue!17}6.2
& \cellcolor{softblue!32}2.5
& \cellcolor{softblue!30}2.7
& \cellcolor{softblue!38}1.9
& \cellcolor{softblue!39}1.8
& \cellcolor{softblue!22}4.4 \\

\makecell[l]{ResNet18\\+Spec}
& \cellcolor{softblue!90}\textbf{100.0}
& \cellcolor{softblue!90}\textbf{100.0}
& \cellcolor{softblue!90}\textbf{100.0}
& \cellcolor{softblue!90}\textbf{0.0}
& \cellcolor{softblue!90}\textbf{0.0}
& \cellcolor{softblue!78}0.3
& \cellcolor{softblue!90}\textbf{0.0}
& \cellcolor{softblue!90}\textbf{0.0}
& \cellcolor{softblue!90}\textbf{0.0}
& \cellcolor{softblue!90}\textbf{0.0}
& \cellcolor{softblue!90}\textbf{0.0}
& \cellcolor{softblue!90}\textbf{0.0}
& \cellcolor{softblue!90}\textbf{0.0} \\

AST
& \cellcolor{softblue!90}\textbf{100.0}
& \cellcolor{softblue!90}\textbf{100.0}
& \cellcolor{softblue!90}\textbf{100.0}
& \cellcolor{softblue!90}\textbf{0.0}
& \cellcolor{softblue!90}\textbf{0.0}
& \cellcolor{softblue!82}0.2
& \cellcolor{softblue!90}\textbf{0.0}
& \cellcolor{softblue!90}\textbf{0.0}
& \cellcolor{softblue!90}\textbf{0.0}
& \cellcolor{softblue!90}\textbf{0.0}
& \cellcolor{softblue!90}\textbf{0.0}
& \cellcolor{softblue!90}\textbf{0.0}
& \cellcolor{softblue!90}\textbf{0.0} \\

SpecTTTra
& \cellcolor{softblue!82}99.8
& \cellcolor{softblue!90}\textbf{100.0}
& \cellcolor{softblue!82}99.8
& \cellcolor{softblue!82}0.2
& \cellcolor{softblue!82}0.2
& \cellcolor{softblue!86}0.1
& \cellcolor{softblue!90}\textbf{0.0}
& \cellcolor{softblue!74}0.4
& \cellcolor{softblue!82}0.2
& \cellcolor{softblue!90}\textbf{0.0}
& \cellcolor{softblue!86}0.1
& \cellcolor{softblue!86}0.1
& \cellcolor{softblue!70}0.5 \\

ConvNeXt
& \cellcolor{softblue!86}99.9
& \cellcolor{softblue!90}\textbf{100.0}
& \cellcolor{softblue!86}99.9
& \cellcolor{softblue!86}0.1
& \cellcolor{softblue!86}0.1
& \cellcolor{softblue!78}0.3
& \cellcolor{softblue!90}\textbf{0.0}
& \cellcolor{softblue!90}\textbf{0.0}
& \cellcolor{softblue!86}0.1
& \cellcolor{softblue!90}\textbf{0.0}
& \cellcolor{softblue!82}0.2
& \cellcolor{softblue!78}0.3
& \cellcolor{softblue!82}0.2 \\

ViT
& \cellcolor{softblue!82}99.8
& \cellcolor{softblue!90}\textbf{100.0}
& \cellcolor{softblue!86}99.9
& \cellcolor{softblue!82}0.2
& \cellcolor{softblue!86}0.1
& \cellcolor{softblue!90}\textbf{0.0}
& \cellcolor{softblue!90}\textbf{0.0}
& \cellcolor{softblue!82}0.2
& \cellcolor{softblue!86}0.1
& \cellcolor{softblue!86}0.1
& \cellcolor{softblue!86}0.1
& \cellcolor{softblue!90}\textbf{0.0}
& \cellcolor{softblue!82}0.2 \\

\bottomrule
\end{NiceTabular}
}
\endgroup
\end{table*}

Table~\ref{tab:30s_results} shows the performance of detectors on 30s music segments. We observe that longer duration generally improves accuracy across all models. We exclude JASCO and Mustango from this evaluation because their outputs are typically shorter than 30 seconds, making direct comparison with other generators infeasible.

\subsection{Inference Time Analysis}
\label{subsec:inference_time}
Table~\ref{tab:inference_time} reports per-track inference latency under a standardized runtime setup.
All models are evaluated with a batch size of 32 and eight data-loading workers on a single NVIDIA H100 GPU.
For context, Table~\ref{tab:detectors} provides corresponding parameter counts and model sizes, enabling interpretation of speed–accuracy trade-offs.

\begin{table}[H]
\centering
\caption{Per-track inference time comparison.}
\label{tab:inference_time}

\setlength{\tabcolsep}{6pt}
\renewcommand{\arraystretch}{1.1}
\rowcolors{2}{white}{lightblue}

\resizebox{0.33\columnwidth}{!}{%
\begin{tabular}{l c}
\rowcolor{darkblue}
\color{white}\textbf{Automatic Detector} &
\color{white}\textbf{Time (ms)} \\
Q-SVM     & 0.069 \\
Deezer & 0.063 \\
M5                  & 0.20 \\
RawNet2             & 2.55 \\
ResNet18{+}Spec     & 1.11 \\
AST                 & 4.25 \\
SpecTTTra           & 0.28 \\
ConvNeXt            & 0.35 \\
ViT                 & 0.17 \\

\bottomrule
\end{tabular}}
\end{table}
% \begin{table}[t]
% \centering
% \caption{Per-track inference time comparison.}
% \vspace{-2mm}
% \label{tab:inference_time}

% \setlength{\tabcolsep}{6pt}
% \renewcommand{\arraystretch}{1.1}
% \rowcolors{2}{white}{lightblue}

% \resizebox{0.43\columnwidth}{!}{%
% \begin{tabular}{l c}
% \rowcolor{darkblue}
% \color{white}\textbf{Automatic Detector} &
% \color{white}\textbf{Time (ms)} \\
% Q-SVM     & 0.069 \\
% M5                  & 0.20 \\
% RawNet2             & 2.55 \\
% ResNet18{+}Spec     & 1.11 \\
% AST                 & 4.25 \\
% SpecTTTra           & 0.28 \\
% ConvNeXt            & 0.35 \\
% ViT                 & 0.17 \\
% \bottomrule
% \end{tabular}}
% \vspace{-3mm}
% \end{table}
\clearpage
\section{Acoustic and Generative Artifacts in AI-Generated Music}
\label{sec: appendix-artifacts}

\begin{wrapfigure}[19]{r}{0.46\linewidth}
  \vspace{-4mm}
  \centering
  \includegraphics[width=\linewidth]{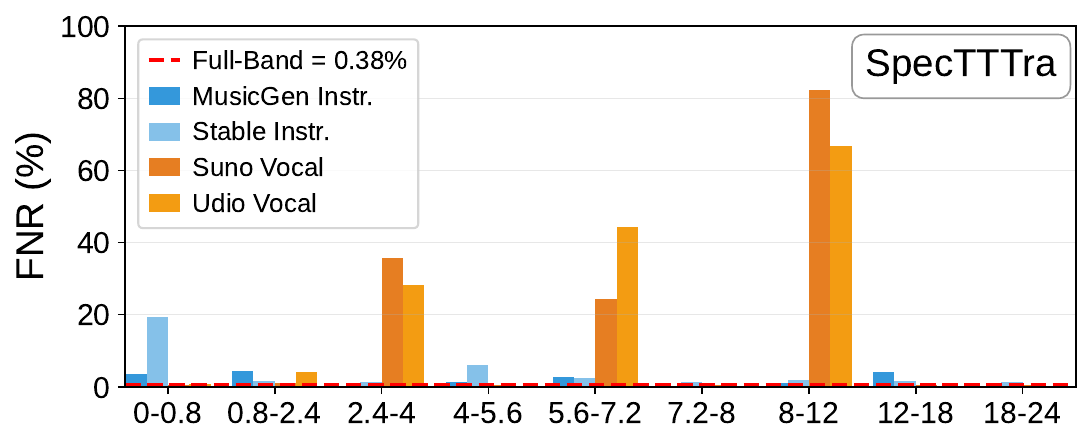}
  \vspace{-2mm}
  \includegraphics[width=\linewidth]{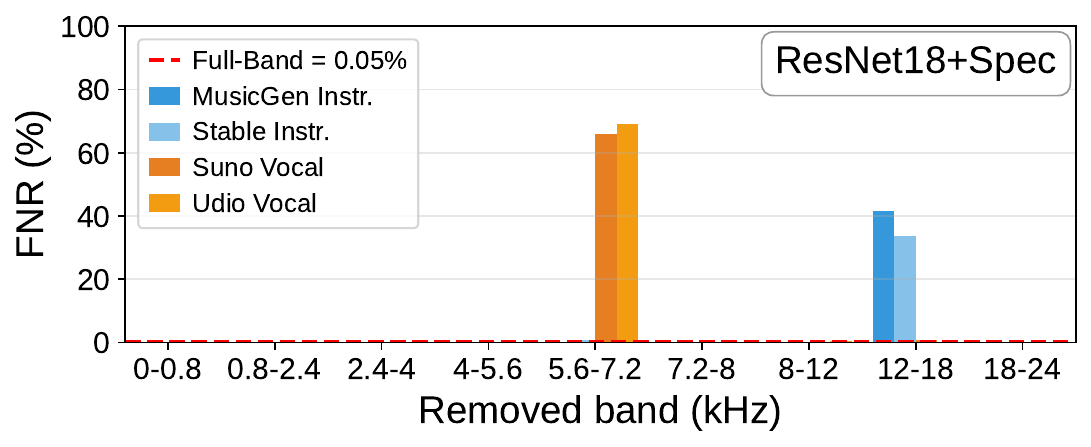}
  \vspace{-2mm}
  \caption{FNR of detectors when removing specific sub-frequency bands for AI-generated music. ``Full-band'' indicates the unfiltered baseline for comparison.}
  \label{fig:bandstop}
\end{wrapfigure}
\vspace{-2mm}
\subsection{Frequency-Sensitivity Analysis}
\label{subsubsec:freq}
To understand which frequency regions most influence AI-music detection, we perform a frequency-sensitivity analysis by selectively suppressing narrow spectral bands and observing how detector performance changes, which converts potential spectral artifacts into measurable evidence of vulnerability. Specifically, a 4th-order Butterworth band-stop filter is applied to one sub-band at a time, and each model's output is compared with its full-band baseline across four sets of AI-generated music in \BenchName, including MusicGen and Stable Audio Open (\textit{instrumental}) and Suno and Udio (\textit{vocal}), using seven detectors. Results for SpecTTTra and ResNet18+Spec are shown in Figure~\ref{fig:bandstop}, with others shown in Figure~\ref{fig:bandstops_appendix}. 

Figure~\ref{fig:bandstops_appendix} shows the impact of band-stop ablations on FNR for M5, RawNet2, ConvNeXt, ViT, and AST. Peaks mark the frequency regions each detector relies on: spectrogram backbones (ConvNeXt/ViT) are most sensitive around 8-12 kHz on vocal sets; AST shows additional reliance in 12-18 kHz for instrumental material; and waveform models (M5/RawNet2) exhibit greater sensitivity in the low bands.

Across \textit{vocal} datasets, we find that detectors do not rely uniformly on the full spectrum but concentrate on several distinctive ``artifact bands.'' For instance, SpecTTTra shows high sensitivity in 2.4--4, 5.6--7.2, and 8--12 kHz, while ResNet18+Spec is particularly fragile around 5.6--7.2 kHz: removing these bands shorply increases their false negatives. For \textit{instrumental} music, SpecTTTra and ResNet18+Spec are most sensitive within 0--0.8 kHz and 12--18 kHz, respectively. As all models share the same training data, these differences likely arise from architectural biases that shape how each model encodes spectral evidence. This observation highlights that detectors rely on narrow, model-specific cues rather than holistic representations, suggesting future designs could improve robustness by promoting cross-band learning, adaptive spectral weighting, or targeted augmentation in these critical regions.
\vspace{-3mm}
\begin{wrapfigure}[22]{r}{0.45\linewidth}
    \centering
    \includegraphics[width=\linewidth]{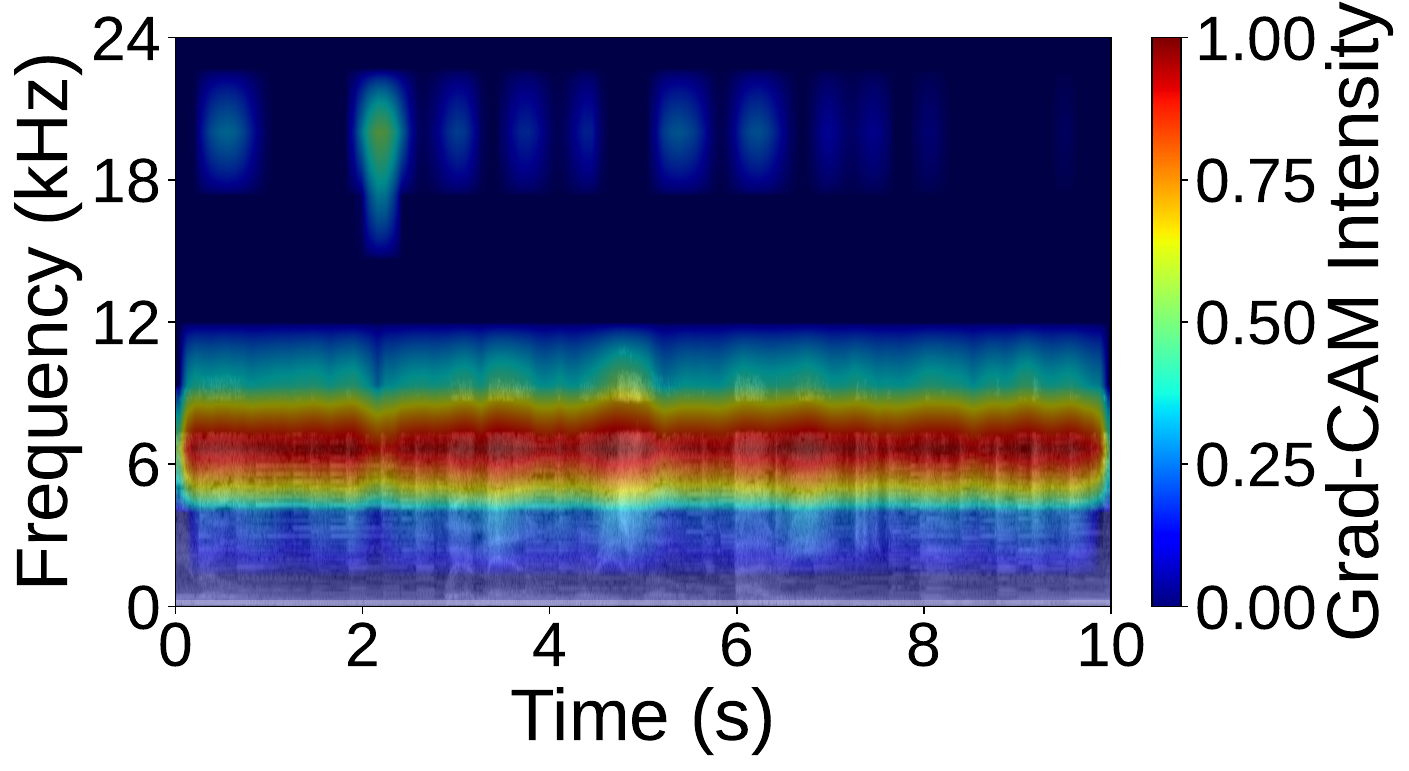}
    \includegraphics[width=\linewidth]{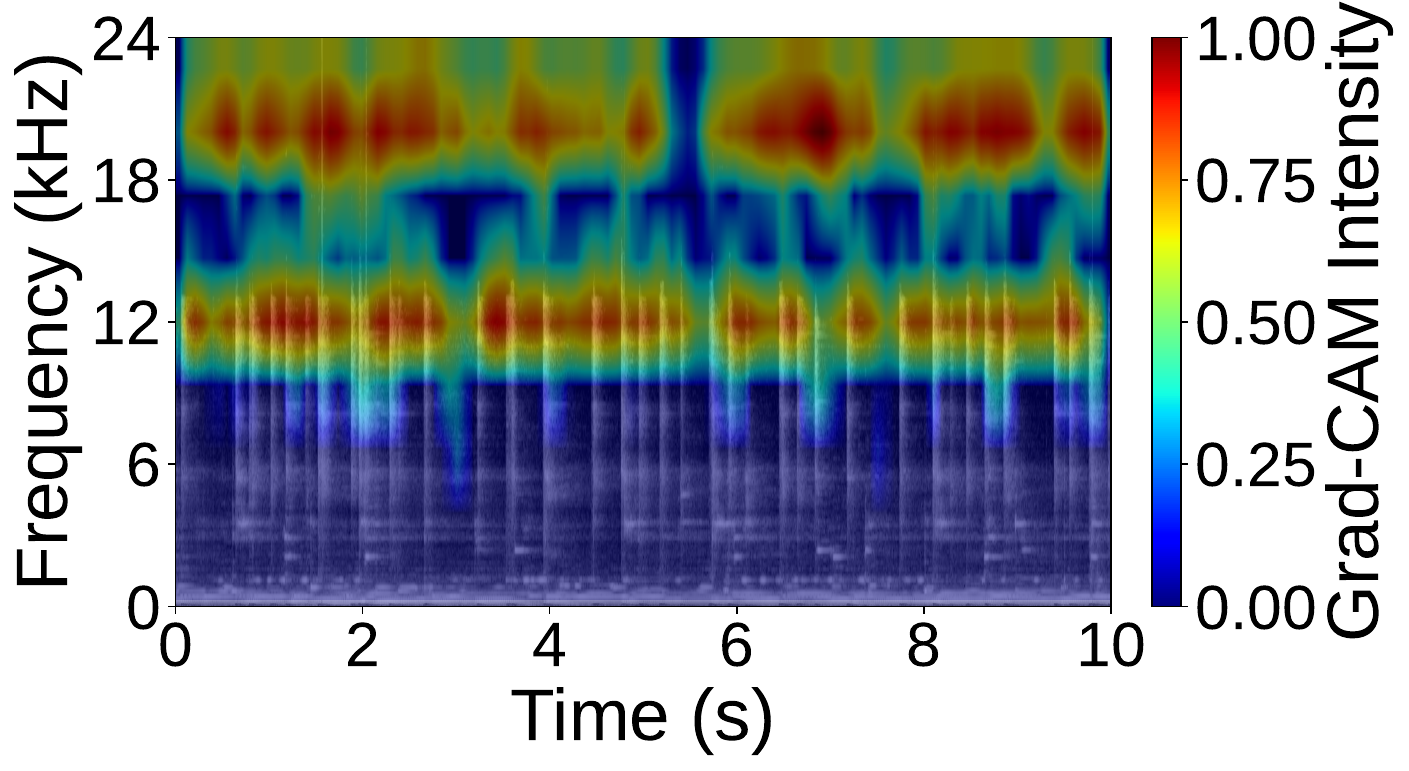}
    \caption{Grad-CAM saliency of ResNet18+Spec on log-mel spectrograms: Udio (top) and Stable Audio Open (bottom).}
    \label{fig:gradcam}
\end{wrapfigure}

To visualize how detectors attend to different spectral regions, we apply Grad-CAM analysis~\cite{selvaraju2017grad} on ResNet18+Spec using representative tracks from the \textit{vocal} and \textit{instrumental} sets. As shown in Figure~\ref{fig:gradcam}, activations for vocal tracks cluster densely around 5--8 kHz, consistent with the most sensitive band identified earlier. In typical singing, even when extended to the whistle register~\cite{WikipediaWhistleRegister}, this region rarely contains the vocal-fold fundamental. Energy within 5--8 kHz primarily arises from harmonic overtones~\cite{titze2021effect} and sibilant or fricative noise~\cite{Pont2019CAA_SibilantS}, both of which AI generators often reproduce imperfectly due to vocoder or phase-reconstruction limitations~\cite{Zhu2024MusicHiFi,Marafioti2019ContextEncoder,LeRoux2008SAPA}. These factors make the 5--8 kHz band a persistent source of detectable artifacts in AI-generated singing.

For instrumental tracks, activations are concentrated in the 12–18 kHz band. AI generators can produce atypical sound in this band when synthesizing instruments that should not naturally have significant energy in this frequency range. Legitimate content includes high-frequency and sometimes ultrasonic energy from metal percussion such as cymbals~\cite{Petrosino2016,brezas2024vibrational} and high-order harmonics from other instruments rather than their fundamentals~\cite{UNSW_Harmonics}. Because most instruments do not place fundamentals in this band and only specific attacks or upper partials occupy it, abnormal or persistent salience at 12–18 kHz provides a clear cue for detectors to distinguish AI-generated music.
\begin{figure}[t]
  \centering
  \includegraphics[width=0.65\linewidth]{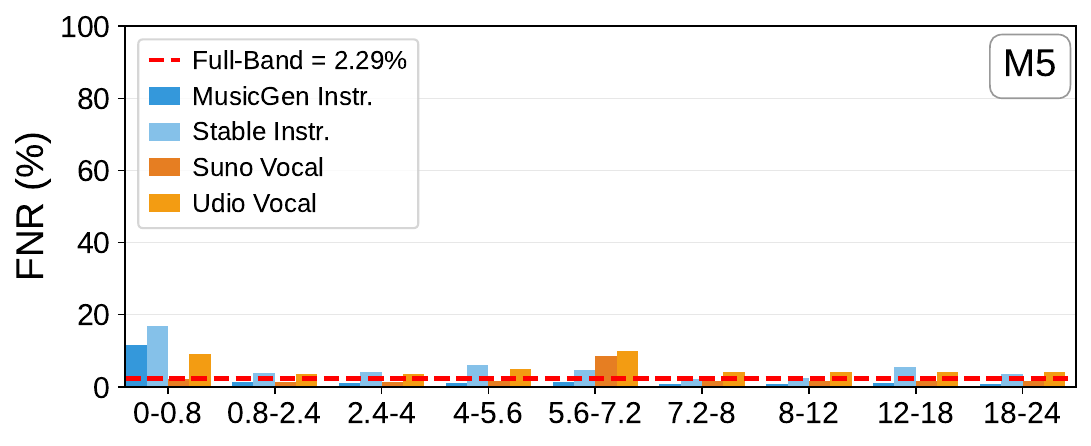}
  \vspace{-1mm}
  \includegraphics[width=0.65\linewidth]{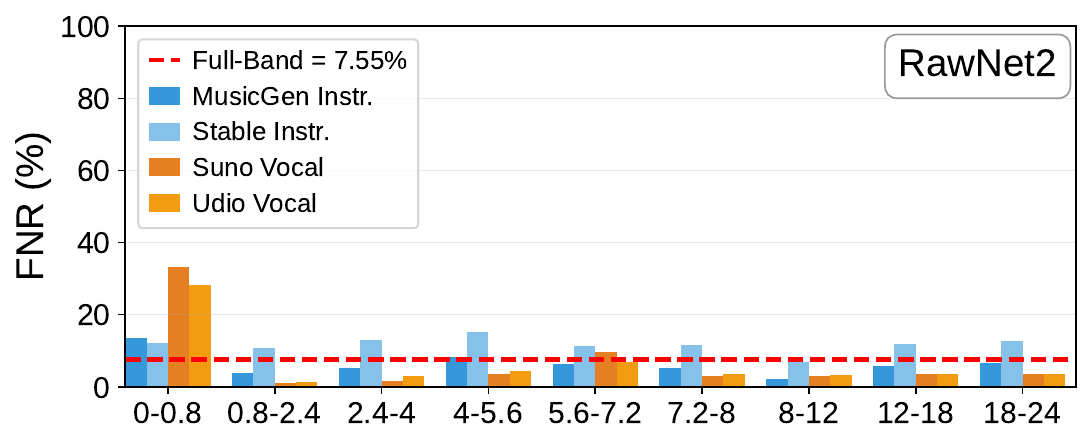}
  \vspace{-1mm}
  \includegraphics[width=0.65\linewidth]{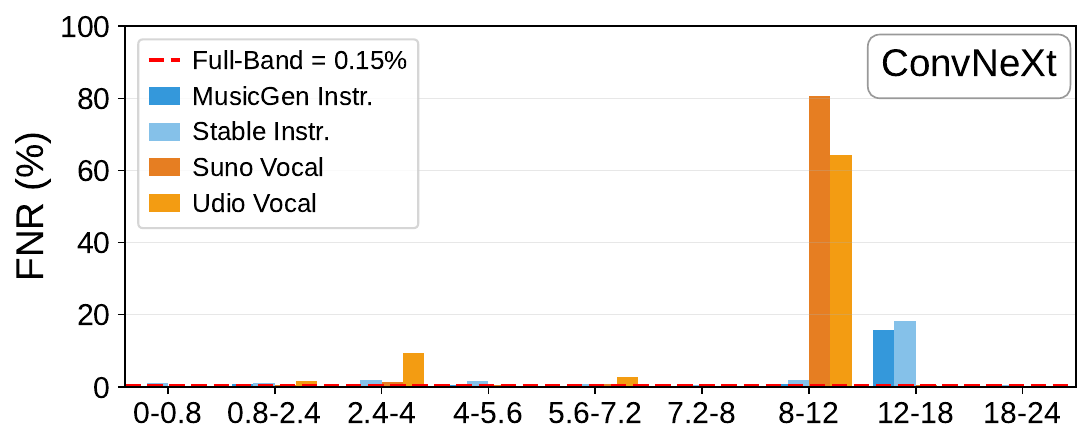}
  \vspace{-1mm}
  \includegraphics[width=0.65\linewidth]{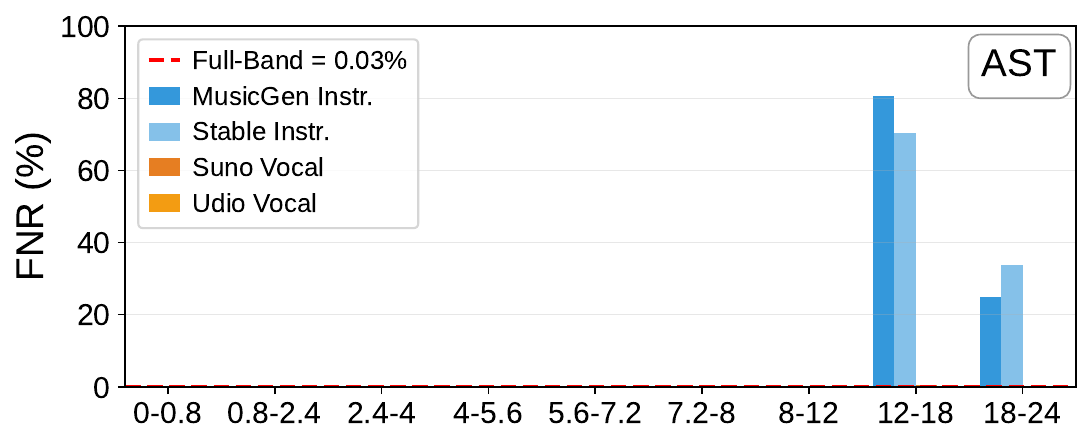}
  \vspace{-1mm}
  \includegraphics[width=0.65\linewidth]{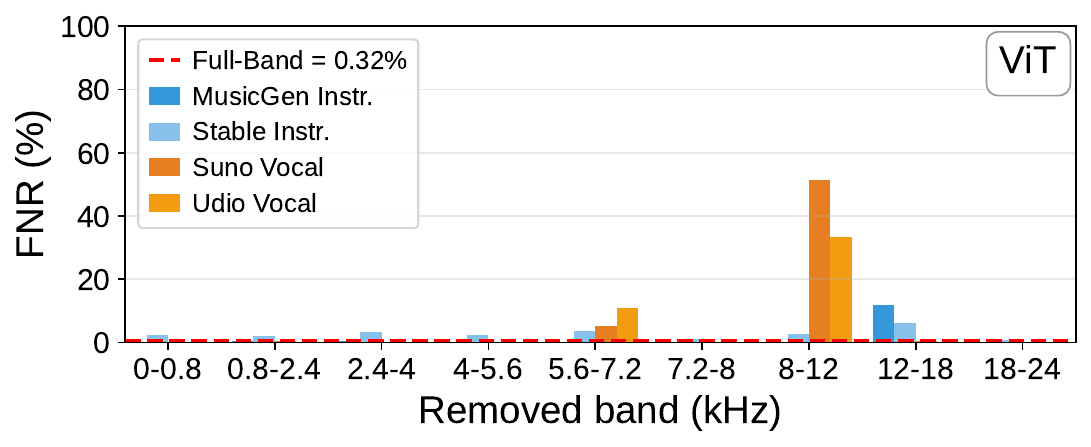}
  \vspace{-2mm}
  \caption{False-negative rate of detectors when removing specific sub-frequency bands for AI-generated music.}
  \label{fig:bandstops_appendix}
  \vspace{-6mm}
\end{figure}

\subsection{Deconvolution Artifacts and Second-Order Spectral Analysis}

Artifacts from transposed convolutions, a.k.a., deconvolution or checkerboard artifacts, were first observed in image synthesis~\cite{odena2016checkerboard} and later found to produce analogous spectral irregularities in audio~\cite{donahue2019wavegan,pons2021upsampling,afchar2025fourier}. These arise when the kernel size and stride interact unevenly, creating periodic amplitude modulations that manifest as subtle spectral line patterns. Many music generation models, including those using the HiFi-GAN~\cite{kong2020hifigan} or neural compressors such as EnCodec~\cite{defossez2022encodec} (e.g., MusicGen, MAGNeT), inherently perform such transposed convolution upsampling, leaving similar structural fingerprints in their outputs. To further analyze this effect, we probed EnCodec with a 1s unit-impulse (details in Appendix~\ref{sec:encodec}), whose reconstruction reveals clear periodic amplitude modulations.
\begin{figure}[H]
  \centering
  \begin{subfigure}[t]{0.42\linewidth}
    \centering
    \includegraphics[width=\linewidth]{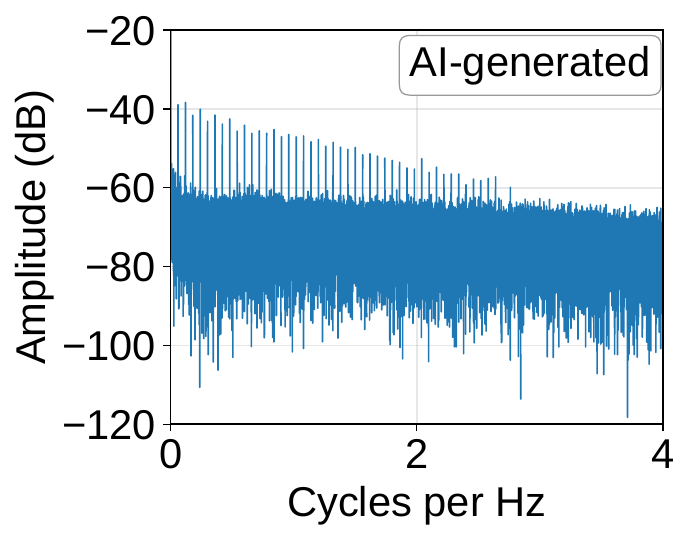}
    \vspace{-3mm}
  \end{subfigure}\hfill
  \begin{subfigure}[t]{0.42\linewidth}
    \centering
    \includegraphics[width=\linewidth]{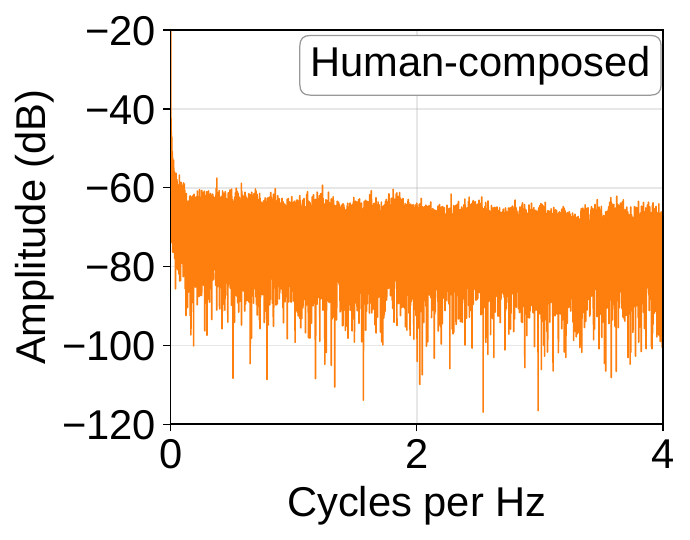}
    \vspace{-4mm}
  \end{subfigure}
  \vspace{-4mm}
  \caption{Second-order FFT of 10s clips comparing AI-generated (JASCO) and human-composed music.} 
  \vspace{-4mm}
  \label{fig:secondfft}
\end{figure}

\vspace{-3mm}
\subsection{Additional Deconvolution Artifact Analysis}
\label{sec:encodec}
We probe EnCodec's 48 kHz model (24 kbps bandwidth) using a 1 s stereo unit impulse passed through the standard encode–decode pipeline. The decoded waveform's FFT is converted to decibel magnitudes, peak-normalized to 0 dB, and plotted over 0–24 kHz. As shown in Figure~\ref{fig: impulse}, the spectrum exhibits periodic amplitude ripples absent from the flat input, consistent with artifacts from transposed-convolution upsampling.

Although prior work has hinted at their presence, such periodic patterns are often masked by rich musical content and remain difficult to isolate using standard FFT magnitude spectra. In our analysis, we find that applying a second-order FFT, i.e., the FFT of the magnitude spectrum, reveals these artifacts far more distinctly, as shown in Figure~\ref{fig:secondfft}. This representation captures repetition frequency along the x-axis and its strength along the y-axis, making periodic structures induced by upsampling appear as clean, evenly spaced spikes. In contrast, human performances exhibit diffuse, irregular energy distributions driven by natural pitch and timbre variation.
This analysis highlights the potential of second-order spectral representations as a diagnostic tool for detecting and interpreting AI-induced artifacts, offering promise for improved detector interpretability, model fingerprinting, and artifact-aware training.

\begin{figure}[H]
  \centering
  \includegraphics[width=0.75\linewidth]{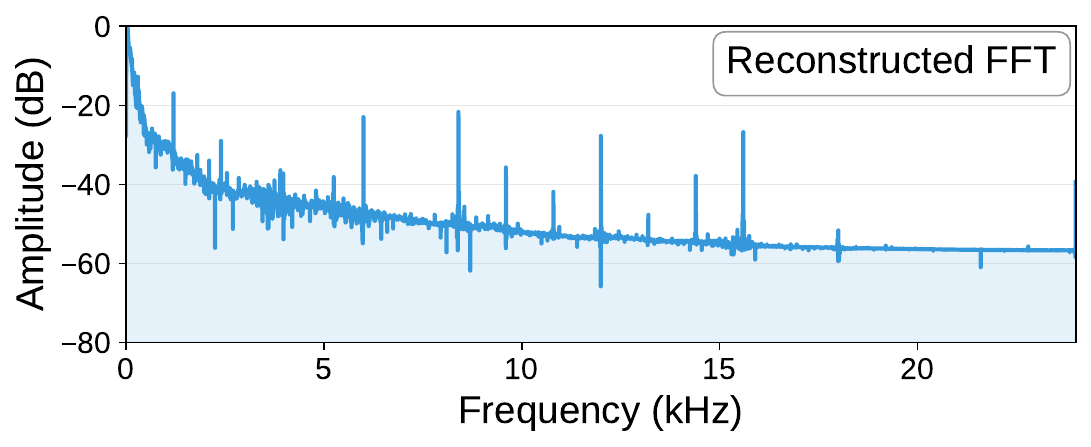}
  \caption{EnCodec impulse-response spectrum.}
  \label{fig: impulse}
\end{figure}
\subsection{Chord Progression Artifacts}
\label{sec:chord}
Another distinctive artifact appears in the form of unnatural or stylistically inconsistent chord progression sequences that violate the grammar of genre and musical tradition. Human composers, especially in genres like jazz, pop, and rock, follow well-established harmonic conventions that shape musical tension and resolution. By contrast, many AI-generated tracks 
exhibit abrupt key shifts, non-diatonic insertions, or loops with reduced cadence density and tonal-tension variability, leading to harmonically incoherent passages.

We analyze chord sequences extracted from 30 AI-generated rock tracks produced by Udio and Suno using the \texttt{chord-extractor} library\footnote{https://github.com/ohollo/chord-extractor.}. Most samples deviate substantially from genre-consistent harmonic patterns\footnote{https://www.guitartricks.com/blog/Rock-Guitar-Chord-Systems.}\footnote{https://blog.native-instruments.com/rock-chord-progressions/.}, frequently inserting isolated non-diatonic tones into otherwise diatonic progressions. To illustrate deviations from conventional tonal logic, Figure~\ref{fig:human-vs-ai} presents two AI-generated chord progressions exhibiting disrupted harmonic resolution.

\noindent\textbf{Progression 1.}
The fourth chord contains a harmonic error that disrupts the natural progression. 
The descending Phrygian sequence is expected to resolve back to C$\sharp$ minor (the tonic of the scale). 
Instead, the progression moves from D major to E$\flat$ major, introducing pitches (E$\flat$, G$\natural$, and B$\flat$) that do not exist within the key signature of three sharps. 
This abrupt modulation sounds unnatural and dissonant within the established context. 
The subsequent return to D major in the final chord creates an irregular harmonic trajectory, breaking from the natural resolution tendency of the progression.

\noindent\textbf{Progression 2.}
The third, fourth, and fifth chords all contain significant harmonic deviations. 
While the first bar generally follows a typical progression in E minor (\textit{ii–i–iv–v}), the pattern breaks on beat 3, where the root notes of the third and fourth chords (A and B) are flattened to A$\flat$ and B$\flat$, respectively. 
This semitone shift introduces strong dissonance and abruptly departs from the established key. 
The final chord, G$\flat$ major, though not inherently dissonant, establishes a tonal center nearly opposite to E minor on the circle of fifths. 
Consequently, the progression fails to provide logical harmonic closure, leaving the listener without a sense of rest or resolution.
\vspace{-3mm}

\begin{figure}[H]
  \centering

  % ===== Row 1 as one subfigure (two images inside) =====
  \begin{subfigure}[t]{\linewidth}
    \centering
    \begin{minipage}{0.48\linewidth}
      \centering
      \includegraphics[width=\linewidth]{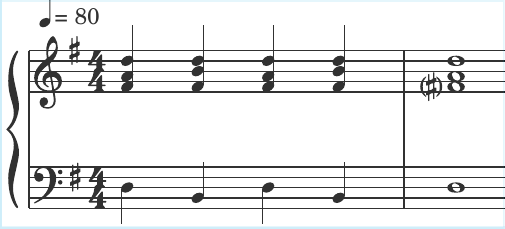}
    \end{minipage}\hfill
    \begin{minipage}{0.48\linewidth}
      \centering
      \includegraphics[width=\linewidth]{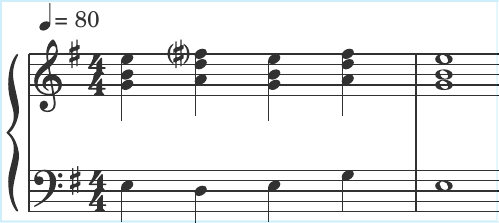}
    \end{minipage}
    \caption{Chord progression of human-composed music}
    \label{fig:rowA}
  \end{subfigure}

  \vspace{0.8em}
    \begin{subfigure}[t]{\linewidth}
    \centering
    \begin{minipage}{0.48\linewidth}
      \centering
      \includegraphics[width=\linewidth]{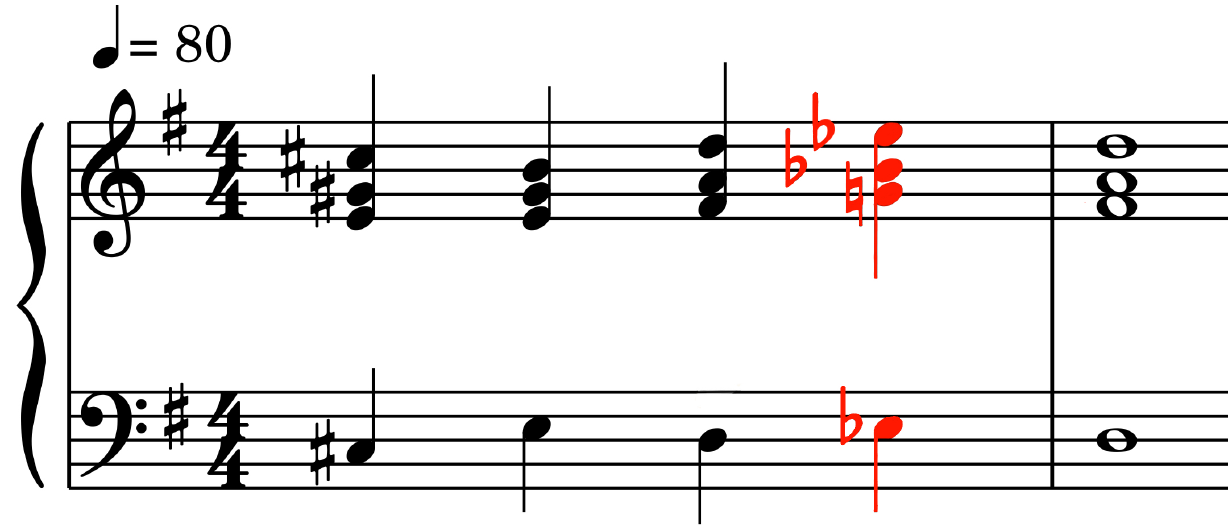}
    \end{minipage}
    \hfill
    \begin{minipage}{0.48\linewidth}
      \centering
      \includegraphics[width=\linewidth]{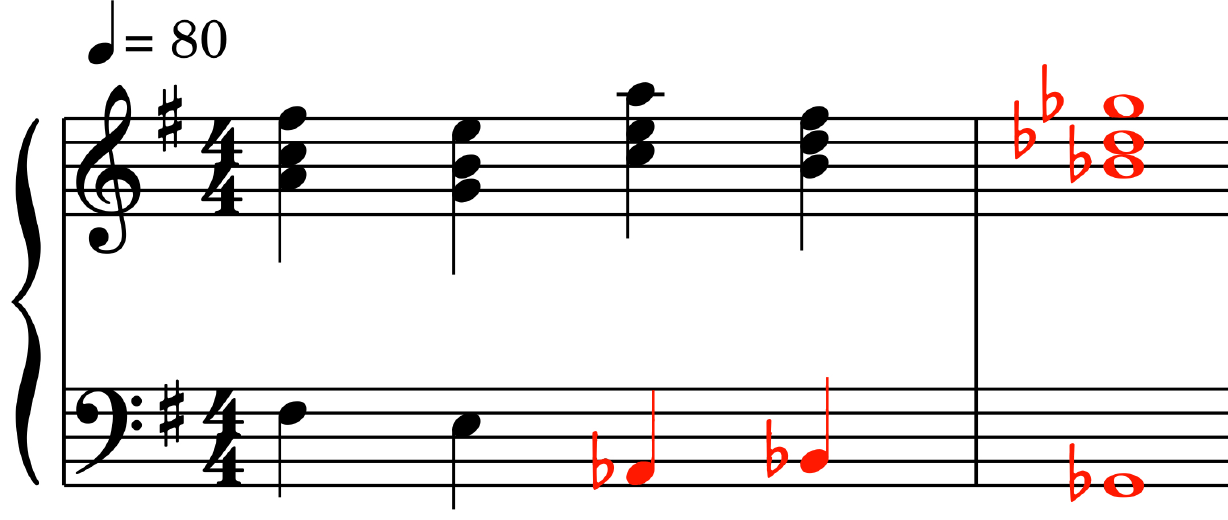}
    \end{minipage}
    \vspace{-2mm}
    \caption{Chord progression of AI-generated music}
    \label{fig:rowB}
  \end{subfigure}
  \caption{Human vs. AI chord progression. Each row shows two 2-bar grand-staff examples.}
  \label{fig:human-vs-ai}
\end{figure}

\vspace{-1mm}
\subsection{Performance-Constraint Artifacts}
\label{sec:performance-constraint}
We also identify artifacts that violate the physical limits of human performance and instrument design. AI-generated pieces occasionally include events that are mechanically impossible, such as piano passages exceeding ten simultaneous notes, guitar phrases extending beyond the instrument's fretboard range, or wind lines that disregard breathing constraints. Such violations rarely occur in human compositions, where anatomical and mechanical boundaries implicitly shape musical structure.
\setlength{\intextsep}{4pt}
\setlength{\columnsep}{8pt}

These inconsistencies stem from the symbolic or spectral nature of generative models, which lack explicit representations of instrument mechanics or performer physiology. Training data capture note co-occurrences but not feasibility constraints, leading models to optimize for statistical plausibility rather than physical realizability.
As shown in Figure~\ref{fig:note violation}, AI-generated guitar tracks contain notes outside the playable range of E2--E6 (MIDI 40--88), while piano sequences exhibit extreme polyphony exceeding human hand span. Although these passages may sound coherent, they reveal structurally implausible gestures that distinguish AI-generated music from human performance.

\begin{figure}[H]
    \centering
    % First subfigure
    \begin{subfigure}[t]{0.44\columnwidth}
        \centering
        \includegraphics[width=\linewidth]{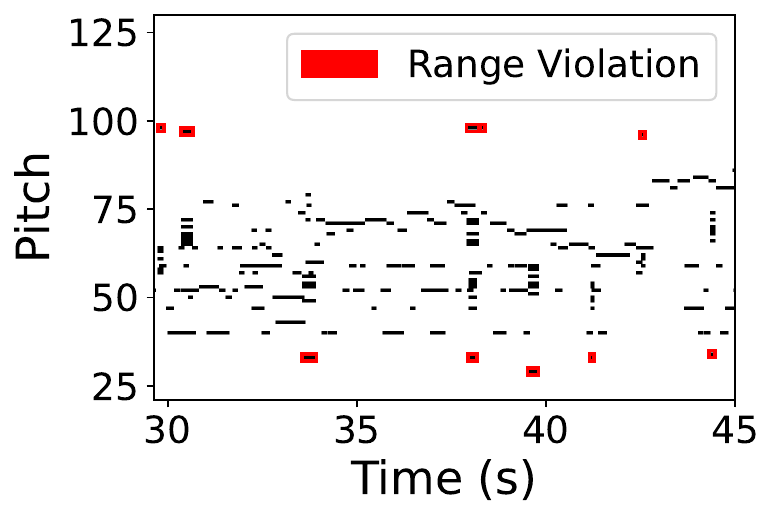}
        \vspace{-6mm}
        \caption{AI-generated guitar notes}
        \label{fig:guitar}
    \end{subfigure}
    \hfill
    % Second subfigure
    \begin{subfigure}[t]{0.44\columnwidth}
        \centering
        \includegraphics[width=\linewidth]{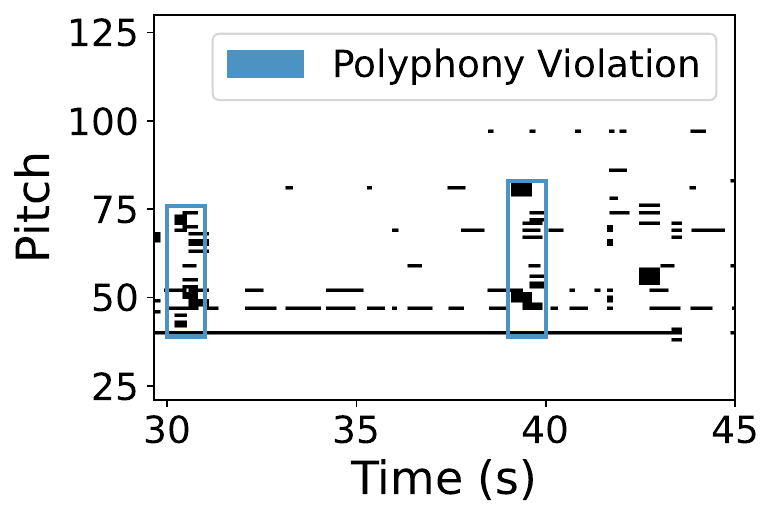}
        \vspace{-6mm}
        \caption{AI-generated piano notes}
        \label{fig:piano}
    \end{subfigure}
    \caption{Examples of performance-constraint artifacts in AI generated music.}
    \label{fig:note violation}
\end{figure}

\subsection{Lyrical Artifacts}
\label{sec:lyrics}
AI-generated singing often exhibits linguistic artifacts that undermine intelligibility and vocal realism. We identify two main types: \textit{pseudo-lyric artifacts} and \textit{approximate-pronunciation artifacts}.

Pseudo-lyric artifacts occur when models generate syllable-like sequences that match musical rhythm but lack semantic meaning. Without explicit lyric conditioning, the output contains speech-like vocalizations such as ``da-lon sa-vah'' or ``me-roh nay''. Even when lyrics are provided, vowel–consonant boundaries often blur, creating the illusion of words without true linguistic content. These errors arise because the model captures phoneme co-occurrence patterns rather than the underlying language-to-phoneme mapping, prioritizing rhythmic smoothness over lexical accuracy.
Additionally, approximate-pronunciation artifacts appear when lyrics are partially reproduced but acoustically distorted. Vowels drift in formant space, consonants weaken, and stress patterns invert. For instance, ``\emph{beautiful}'' becomes ``\emph{byoo-tuh-fuhl}'', and ``\emph{together}'' shifts to ``\emph{TOgether}''. Such artifacts reflect the tension between melodic continuity and articulatory precision: models trained for spectral smoothness often neglect phonetic sharpness. As also noted in Section~\ref{subsubsec:freq}, a related issue is the \textit{sibilance artifact}, where fricatives like ``s'', ``sh'', and ``z'' generate exaggerated hiss between 5--8~kHz (e.g., ``\emph{ships at sea}'' rendered as ``\emph{shh-ips at s\textsuperscript{h}ea}''). These broadband noise bursts arise from vocoder limitations in modeling the stochastic airflow of natural fricatives.

\clearpage

\section{Ethical Considerations}
This study complied with Institutional Review Board (IRB) standards for human-subject research and responsible data use.
All listening studies were reviewed and deemed exempt by the university's IRB.
Participants provided informed consent and were informed that their anonymized responses would be used solely for academic research.
No personally identifiable information was collected beyond optional demographic details such as musical background or listening experience.

\noindent\textbf{Data Sources and Copyright.}
Human-composed music included in \BenchName was drawn exclusively from publicly available, verifiably human-authored datasets (e.g., Genius-Lyrics and JamendoMaxCaps) under terms permitting research use. All AI-generated and co-created materials were produced by the authors using open-source or commercial APIs in compliance with their respective terms of service. The dataset does not include copyrighted works or artist likenesses without authorization, nor does it distribute any protected material. For the human--AI co-creation component, we involved musicians to provide qualitative feedback on musical compositions. Participants provided informed consent and were informed that their anonymized responses would be used solely for academic research. No personally identifiable information was collected beyond optional demographic details, such as musical background or listening experience.

\noindent\textbf{Responsible AI Use.}
The generative models employed (e.g., MusicGen, AudioLDM2, Suno, Udio) were used strictly for evaluation and not for commercial release or reproduction of copyrighted styles or performances. We explicitly avoid generating or disseminating content that could infringe on the rights or reputations of living artists. \BenchName is intended solely for research on detection and interpretability, not for creating derivative musical works.

\vspace{-2mm}

\section{Generative AI Usage}
LLMs were used for limited editorial assistance, such as improving grammar, wording, and clarity in some parts of the paper. All scientific content, such as dataset design, experiments, analyses, and conclusions, was fully developed and validated by the authors. LLMs also supported two text-generation components within our dataset construction pipeline: 
\begin{itemize} [leftmargin=*, itemsep=0pt, topsep=0pt]
\item Caption Consolidation for Prompt Construction (Section~\ref{sec:AI-Generated Music}).
Segment-level captions generated by LP-MusicCaps were merged into coherent track-level descriptions using GPT-OSS following a fixed template. Authors manually verified that all merged captions faithfully reflected the underlying audio and introduced no new musical content.
\item Structuring Musician Feedback for Human–AI Co-Creation (Section~\ref{sec: co-created}).
Free-form comments from professional musicians were reformatted into standardized editing prompts using GPT-OSS. The LLM did not generate musical ideas; it only regularized human-provided feedback. All prompts were inspected to ensure fidelity to the original intent.
\end{itemize}

These uses do not influence the scientific findings of the paper: no experiment, detector evaluation, artifact analysis, or statistical conclusion depends on nondeterministic LLM behavior. All prompts and templates used in these steps are released with MuseBench to ensure reproducibility.

No LLMs were trained for this work, and LLM usage was limited to lightweight text transformation. Environmental impact was therefore minimal. All datasets and audio sources comply with licensing and platform policies.

% \newpage
% \input{checklist.tex}

\end{document}